\documentclass{aastex6}

\usepackage{color}
\usepackage{graphicx}
\newcommand{\myemail}{takashi.tsukagoshi@nao.ac.jp}

\fullcollaborationName{The Friends of AASTeX Collaboration}

\begin{document}


\title{The Flared Gas Structure of the Transitional Disk around Sz~91}


\author{
Takashi Tsukagoshi\altaffilmark{1},
Munetake Momose\altaffilmark{2},
Yoshimi Kitamura\altaffilmark{3},
Masao Saito\altaffilmark{1},
Ryohei Kawabe\altaffilmark{1,4,5},
Sean Andrews\altaffilmark{6},
David Wilner\altaffilmark{6},
Tomoyuki Kudo\altaffilmark{7},
Jun Hashimoto\altaffilmark{8},
Nagayoshi Ohashi\altaffilmark{7},
Motohide Tamura\altaffilmark{5,8}
}


\altaffiltext{1}{National Astronomical Observatory Japan (NAOJ), Osawa 2-21-1, Mitaka, Tokyo 181-8588, Japan; \myemail}
\altaffiltext{2}{College of Science, Ibaraki University, Bunkyo 2-1-1, Mito, Ibaraki, 310-8512, Japan}
\altaffiltext{3}{Institute of Space and Astronautical Science, Japan Aerospace Exploration Agency, Yoshinodai 3-1-1, Sagamihara, Kanagawa 229-8510, Japan}
\altaffiltext{4}{SOKENDAI (The Graduate University for Advanced Studies), 2-21-1 Osawa, Mitaka, Tokyo 181-8588, Japan}
\altaffiltext{5}{Department of Astronomy, School of Science, University of Tokyo, Bunkyo, Tokyo 113-0033, Japan}
\altaffiltext{6}{Harvard-Smithsonian Center for Astrophysics, 60 Garden Street, Cambridge, MA 02138, USA}
\altaffiltext{7}{Subaru Telescope, National Astronomical Observatory of Japan, 650 North A’ohoku Place, Hilo, HI96720, USA}
\altaffiltext{8}{Astrobiology Center of NINS, 2-21-1 Osawa, Mitaka, Tokyo, 181-8588, Japan}


\begin{abstract}
We report $0\farcs14$ resolution observations of the dust continuum at band 7, and the CO(3--2) and HCO$^{+}$(4--3) line emissions toward the transitional disk around \object{Sz 91} with Atacama Large Millimeter/submillimeter Array (ALMA).
The dust disk appears to be an axisymmetric ring, peaking a radius of $\sim$95~au from a Gaussian fit.
The Gaussian fit widths of the dust ring are 24.6 and 23.7~au for the major and the minor axes, respectively, indicating that the dust ring is not geometrically thin.
The gas disk extends out to $\sim$320~au and is also detected in the inner hole of the dust ring.
A twin-line pattern is found in the channel maps of CO, which can be interpreted as the emission from the front and rear of the flared gas disk.
We perform the radiative transfer calculations using RADMC-3D, to check whether the twin-line pattern can be reproduced under the assumption that the flared gas disk has a power-law form for the column density and $T_\mathrm{gas}=T_\mathrm{dust}$.
The thermal Monte Carlo calculation in RADMC-3D shows that the disk temperature has a gradient along the vertical direction beyond the dust ring, as it blocks the stellar radiation, and thus the twin-line pattern can be naturally explained by the flared gas disk in combination with the dust ring.
In addition, no significant depletion of the CO molecules in the cold midplane achieves a reasonable agreement with the observed twin-line pattern.
This result indicates that the CO emission from the rear surface must be heavily absorbed in the cold midplane.
\end{abstract}




\section{Introduction} \label{sec:intro}
It is widely believed that protoplanetary disks are the birthplace of planets.
Their physical structures have an important role in controlling how the planetary systems are formed \citep{bib:kokubo2002}.
The evolutionary stage of a protoplanetary disk can be identified based on the slope index of its spectral energy distribution (SED) \citep[e.g.,][]{bib:lada1984,bib:lada1987}.
The excess emission at infrared (IR) wavelengths indicates the existence of extended disk material around the central star.
Transitional disks have been identified as young stellar objects with little or no near-IR (NIR) excesses and significant mid- and/or far-IR excesses in their SEDs \citep{bib:strom1989}.
This peculiar feature of the SEDs implies the clearing or depletion of dust grains in the inner part of their circumstellar disks, i.e., an inner hole.
Such a transitional disk object is believed to be a good laboratory for the investigation of disk evolution from an optically thick primordial disk to an evolved optically thin disk, as well as the planet formation process.\par

Many candidates for the transitional disk objects have been identified in nearby star-forming regions based on their IR spectral indexes.
The complicated SED shape of a transitional disk is described by a two-parameter scheme based on the longest wavelength at which the observed flux density is dominated by the stellar photosphere and the slope of the IR excess measured from which to 24 $\micron$ \citep{bib:cieza2010,bib:romero2012,bib:cieza2012}.
Although the SED-based identification is indirect evidence of the inner hole structure, the dust-clearing/depleted region has been directly resolved by high-resolution observations with (sub-)millimeter interferometers \citep[e.g.,][]{bib:brown2009,bib:andrews2011}.
Recently, Atacama Large Millimeter/submillimetter Array (ALMA) has been unveiling more detailed structures of the transitional disks around low-mass stars, e.g., [PZ99]~J160421.7-213028 \citep{bib:zhang2014,bib:dong2017}, TW~Hya \citep{bib:andrews2016,bib:tsukagoshi2016}, and SR~24S \citep{bib:pinilla2017} and intermediate mass stars, e.g., HD~142527 \citep{bib:casassus2013,bib:fukagawa2013}, IRS~48 \citep{bib:marel2013}, SR~21 and SAO~206462 \citep{bib:perez2014}, and HD~169142\citep{bib:fedele2017}.
Such surveys of the transitional disks with ALMA have revealed that gas typically exists inside the dust cavity, while there is a gas density drop at a smaller radius than the inner radius of the dust cavity \citep{bib:marel2015,bib:marel2016}.
However, the structures of the transition disks, such as the size of the inner hole, disk mass, and degree of gas depletion, are rich in variety.\par

Recent high-resolution and high-sensitivity observations with ALMA have revealed the vertical structure of the gas component in protoplanetary disks \citep{bib:gregorio-mosalvo2013,bib:pinte2017}.
In some channel maps, the CO emission apparently traces two different emitting regions of the rotating gas disk; the bright emission from the disk surface in the front and the fainter emission in the rear.
To reproduce this velocity pattern, the gas disk must be flared enough to separate the front and rear surfaces, and the gas disk should have a CO depleted region in the cold disk midplane to achieve a reasonable agreement with the observations.
These results indicate that the two-layered velocity pattern seen in the optically thick CO emission is a powerful tool to diagnose the vertical structure of gas disks more directly.
So far, the two-layered velocity pattern has been found for two massive full disk samples, HD~163296 \citep{bib:rosenfeld2013,bib:gregorio-mosalvo2013} and IM~Lup \citep{bib:pinte2017}.
No observational information on the two-layered velocity pattern for low-mass transitional disks has been reported, whereas it has been found recently for the transitional disk around the Herbig Ae star HD~97048 \citep{bib:plas2017a} and HD~34282 \citep{bib:plas2017b}.

\object{Sz~91} is a M1.5 young stellar object possessing a transitional disk located in the Lupus III molecular cloud \citep{bib:romero2012}.
The SED shows no significant NIR excess, with a spectral index from 2 to 24 $\micron$ of $-2$ (i.e., class III) and a steep flux rising between 24 and 70 $\micron$ \citep{bib:romero2012}.
Accordingly, there is a large dip around 20 $\micron$, indicating a large inner hole in the dust disk.
The mass accretion rate onto the star is estimated with the VLT/X-Shooter spectrograph to be $\dot{M}_\mathrm{acc}=10^{-8.92}$--$10^{-8.69}$ $M_\sun$ yr$^{-1}$ \citep{bib:alcala2017}.
The estimates of the stellar mass, the stellar radius and the effective temperature are $M_\ast=$0.47 $M_\sun$, $R_\ast=$1.46 $R_\sun$ and $T_\mathrm{eff}=$3720~K, respectively, with a distance of 200~pc \citep{bib:canovas2015}.
However, most recently, the distance to the source is measured to be 156--160~pc based on the GAIA data release 2 \citep{bib:bailer-jones2018}.
In this paper, we adopt a distance of 158~pc.
We estimate the stellar parameters taking into account the new distance to be 0.48 $M_\sun$, 1.28 $R_\sun$ and 3720~K for $M_\ast$, $R_\ast$ and $T_\mathrm{eff}$, respectively, using the stellar evolutional tracks of \citet{bib:siess2000}.
The stellar age is estimated to be $\sim$3~Myr.
We employ these stellar parameters throughout the following sections.\par

\citet{bib:tsukagoshi2014} have resolved the inner hole structure through high-resolution observations at both submillimeter and NIR wavelengths.
The total flux density at 345 GHz was measured to be 32.1 mJy, corresponding to a gas mass of $\sim10^{-3}$ $M_\sun$.
The authors suggest the existence of the inner hole based on the fact that the peak position of the 345 GHz emissions does not coincide with the stellar position.
The NIR image shows a crescent-like distribution around the central star, and its radius is 65~au.
The NIR emissions trace the inner region of the 345 GHz emission, suggestive of scattered light in the inner most part of the dust disk.
Based on the SED modeling study, the authors have found that the observed SED can be reproduced by the combination of a cold outer disk imaged at 345 GHz and an additional unseen hot component with a temperature of $\sim$180~K in the inner hole.\par

The inner hole structure has also been resolved with recent ALMA observations.
\citet{bib:canovas2015} reported the presence of the largest dust cavity, whose radius is estimated to be 97~au at band 6 (231~GHz) adopting a distance of 200~pc.
The authors also found that the CO gas extends up to $\sim400$~au from the star and also exists in the dust cavity.
The symmetrical dust ring structure has been revealed by higher resolution ($\sim0\farcs2$) ALMA observations \citep{bib:canovas2016}.
The emissions are concentrated at a radius of 110~au, with a width of 44~au.
The authors pointed out that the ring-like distribution of the dust emissions and the different inner radii between the submillimeter continuum and the CO emissions are supposed to be due to the disk-planet interaction because the hydrodynamical simulations including fragmentation, coagulation, and radial drift of dust particles indicate that large (millimeter-sized) grains accumulate at the pressure maximum caused by the planet-disk interaction \citep{bib:pinilla2012}.\par

In this paper, we report the results of new observations of Sz~91 with higher resolutions and better sensitivity than the previous observations.
The improvement of the image quality is essential in order to determine the detailed structure of the Sz~91's disk because the resolution of the previous ALMA observation is comparable to the width of the dust ring and the sensitivity of the observation is insufficient to determine the structure of the gas disk in detail.
In \S \ref{sec:observation}, we describe our observation and data reduction.
The resulting images of the band 7 continuum emission and the CO(3--2) and HCO$^+$(4--3) molecular line emissions are presented in \S \ref{sec:result}.
In \S \ref{sec:discussion}, we discuss the detailed structure of the Sz~91's disk.
In \S \ref{sec:co_disk}, the complex velocity pattern found in the CO channel maps, which is likely to show the emission from the front and rear sides of a flared disk, is discussed using radiative transfer calculations.
In \S \ref{sec:hot_component}, we briefly mention the hot component introduced in \citet{bib:tsukagoshi2014}.
We summarize this paper in \S \ref{sec:summary}.

\section{Observation}\label{sec:observation}
Our observations of Sz~91 with Atacama Large Milimeter/submillimeter Array (ALMA) were conducted on July 20 and 21, 2015.
During the observed period, 42 antennae were operated, and the antenna configuration was C34-7(6), resulting in a maximum baseline length of $\sim$1.6~km.
The observations were conducted by three execution blocks.
The total integration time for the execution blocks was 34~min each, i.e., 1.5~h in total.
The typical precipitable water vapor during the observation period was ranging from 0.78 to 0.87~mm.
The band~7 receiver system was employed to detect the continuum emissions at band~7 (350 GHz) and the $^{12}$CO(3--2) (345.796 GHz) and HCO$^{+}$(4--3) (356.734 GHz) molecular lines.
The four spectral windows (SPWs) of the correlator were used in the Frequency Division Mode.
Two of them covered the rest frequencies of the target lines, in which the bandwidth and the channel spacing were 117.188~MHz and 30.518~kHz, respectively, which correspond to a velocity resolution of $\sim0.05$ km s$^{-1}$.
The other two spectral windows were configured for the continuum emissions with a bandwidth of 1.875~GHz each, resulting in a total bandwidth of 3.75~GHz.
The calibrators of the bandpass character and the absolute flux scale were J1517-2422 and Pallas, respectively.
The complex gain was calibrated by observing J1610-3958 in a cycle of $\sim$5 min.\par

The visibility data were reduced and calibrated using the Common Astronomical Software Application (CASA) package \citep{bib:mcmullin2007}.
The raw data were calibrated with the reduction script provided by ALMA.
After flagging bad data, the calibration tables for the bandpass, complex gain, and flux scaling were made and applied to the data for each SPW separately.
The continuum visibilities were obtained by averaging the line-free channels in each SPW.
The continuum image was reconstructed by the CLEAN algorithm with the Briggs weighting of a robust parameter of 0.
We also employed the multiscale CLEAN with scale parameters of [0, 0.12, 0.36] arcsec to reconstruct extended emissions effectively.
The resolution of the CLEAN image achieved was $0\farcs134\times0\farcs104$, and thus the final CLEAN image was restored by a Gaussian function to be a symmetric beam with an effective resolution of $0\farcs14$ for image visualization.
The rms noise level of the image was measured to be 61.2 $\mu$Jy beam$^{-1}$.\par

The visibilities of the line emissions were obtained by subtracting the continuum visibilities.
The line emissions were imaged using the multiscale CLEAN algorithm with the same scale parameters as for the continuum image.
The velocity step of the image cubes was set to be 0.2 km s$^{-1}$ to achieve a significant signal-to-noise ratio in each channel map.
The CLEAN images were restored to have the same resolution as in the continuum map, i.e., the beamsize of $0\farcs14$.
The rms noise levels of the CLEAN images were 6.0 and 6.7 mJy beam$^{-1}$ for CO and HCO$^{+}$, respectively.

\section{Results}\label{sec:result}
\subsection{Band 7 Continuum Emission}
\begin{figure}
\begin{center}
	\includegraphics[width=\textwidth]{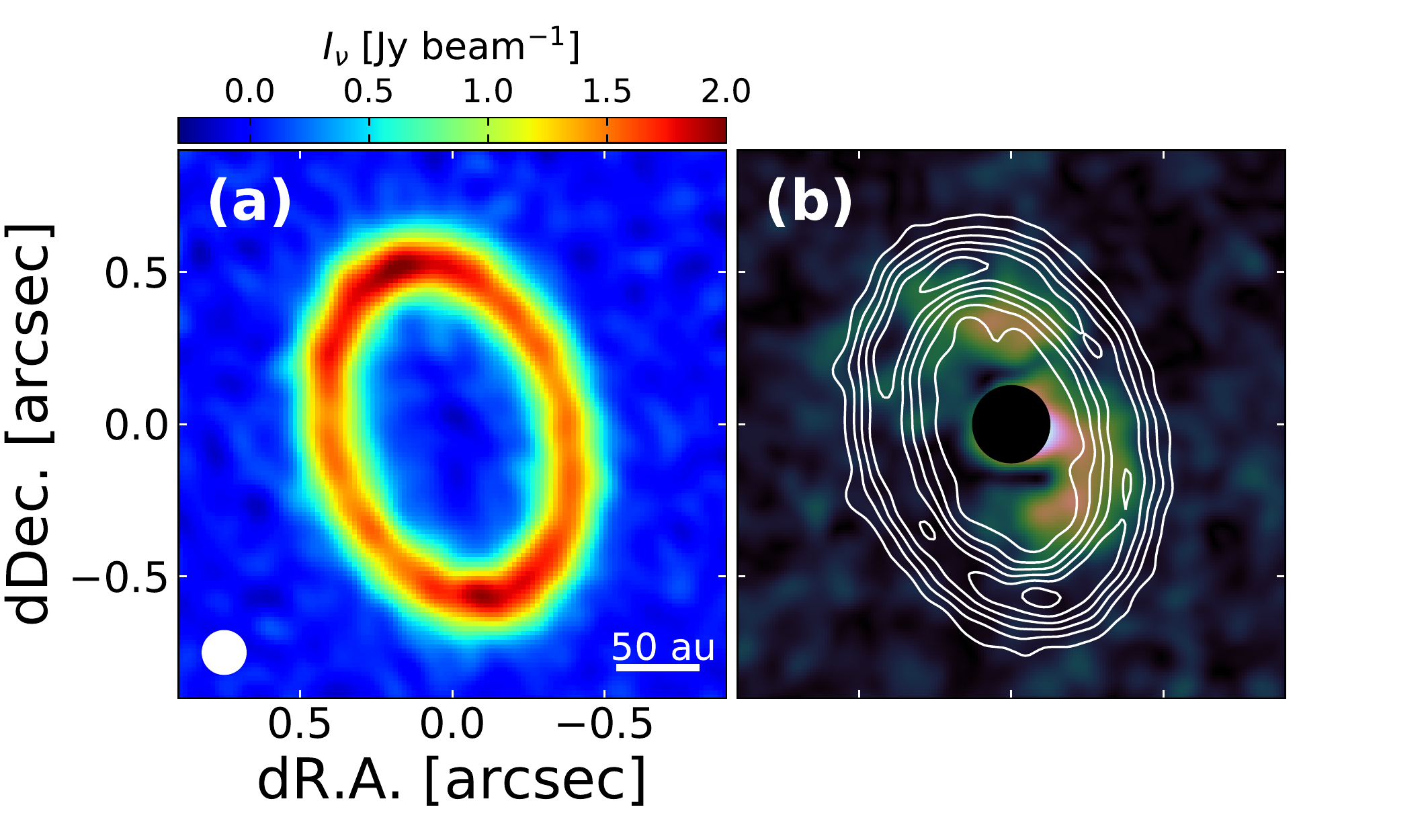}
	\caption{(a) Continuum image at band 7 (350 GHz). The circle at the bottom left corner shows the synthesized beam. (b) Same as (a), but the continuum emission is described with the white contour. The contour starts at 5$\sigma$ with an interval of 5$\sigma$, where 1$\sigma=$61.2 $\mu$Jy beam$^{-1}$. The color scale shows the polarized intensity map for NIR in which the inner area of 0$\farcs$25 diameter (filled circle) is masked \citep{bib:tsukagoshi2014}.}\label{fig:b7cont}
\end{center}
\end{figure}

The CLEANed image of the continuum emission at band 7 is shown in Figure \ref{fig:b7cont}(a).
The emission distributes like a ring and there are intensity peaks along P.A.$=$20$\degr$ and 200$\degr$, i.e., the semi-major axis of the ring.
These features are similar to those found in the previous study \citep{bib:canovas2016}.
Figure \ref{fig:b7cont}(b) also shows the continuum emission at band 7 superposed on the $K_s$ band ($\lambda=2.15\micron$) scattered light image obtained by \citet{bib:tsukagoshi2014}.
The crescent-like $K_s$ emission suggests that this part is the near side of the disk.
The continuum emission clearly surrounds the $K_s$ emission, indicating the radial variation of dust size distribution.
The total flux density above $3\sigma$ is measured to be $45.2\pm0.5$ mJy, which agrees well with that of \citet{bib:canovas2016} in consideration of the difference of the observed frequency.
No clear emission is found in the inner hole of the dust ring above 3$\sigma$ (1$\sigma$=61.2 $\mu$Jy beam$^{-1}$).\par

To measure the size and inclination of the dust ring, we performed a visibility fitting with a ring distribution using the {\it uvfits} task installed on the {\it MIRIAD} software, as \citet{bib:canovas2016} has done.
With this task, we obtained the major and minor axes lengths of $(1\farcs185\pm0\farcs002)\times(0\farcs766\pm0\farcs002)$ and a position angle of $18\fdg1\pm0\fdg2$.
The inclination angle $i$ is estimated from the aspect ratio to be $49\fdg7\pm0\fdg2$.
The position and inclination angles are comparable to but slightly different from those derived in \citet{bib:canovas2016}, probably due to calibration uncertainties.\par

\begin{figure}
\begin{center}
	\includegraphics[width=\textwidth]{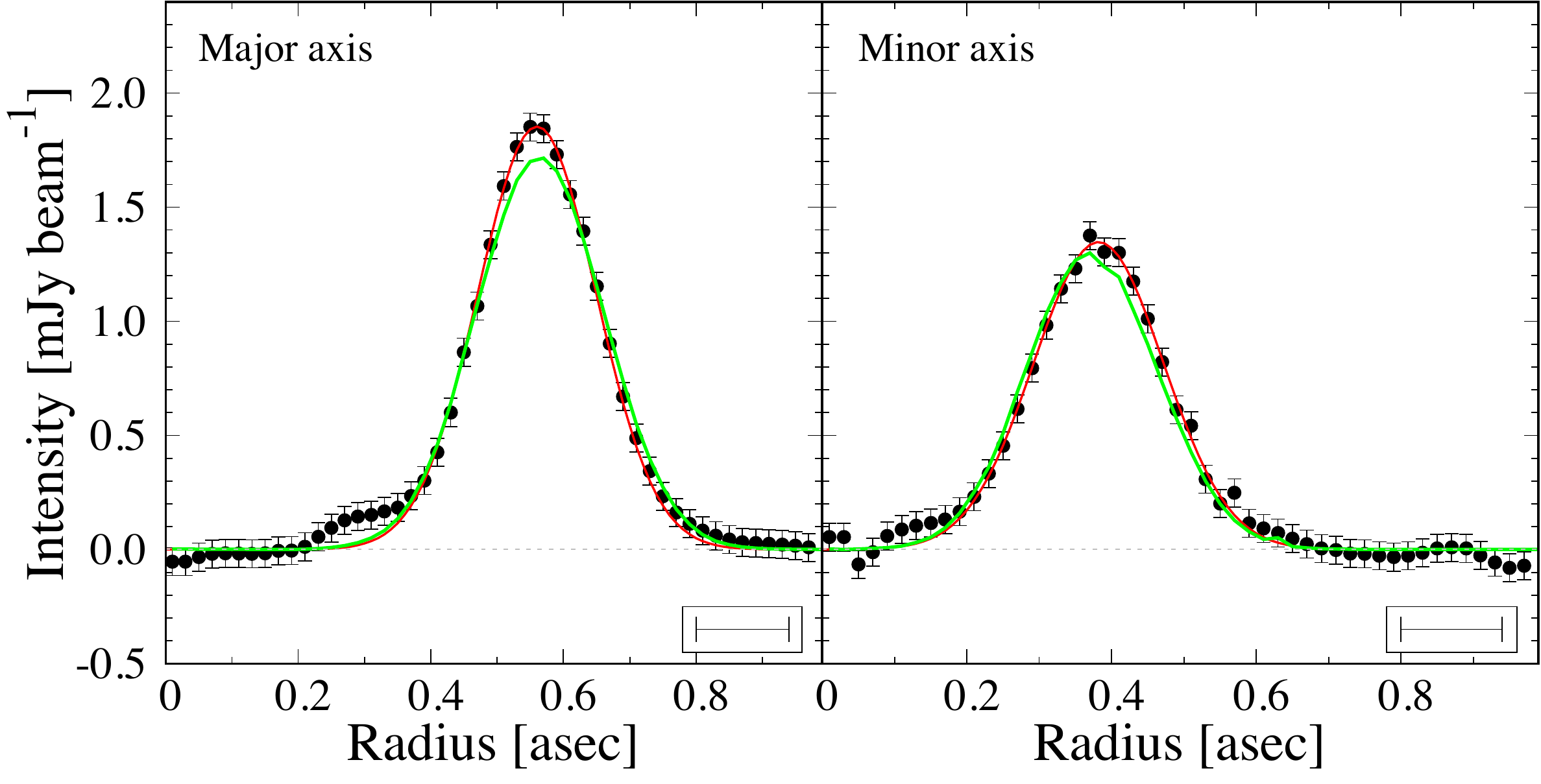}
	\caption{Radial intensity profiles of the dust ring along the major (left panel) and minor (right panel) axes at P.A.$=$18 deg and 108.1 deg, respectively. Image deprojection was not performed. The red line indicates the best-fit curve in Gaussian fitting. The green line shows the radial profile of the model image obtained by the radiative transfer calculation (see \S \ref{sec:co_disk}). The bar at the bottom-right corner shows the FWHM of the synthesized beam ($0\farcs14$).}\label{fig:radplot}
\end{center}
\end{figure}

Figure \ref{fig:radplot} shows the radial intensity profiles of the dust ring along the major and minor axes.
Those widths are determined from the least-squares fitting with a Gaussian function to be 0$\farcs$209$\pm$0$\farcs$001 and 0$\farcs$205$\pm$0$\farcs$001 in terms of the full width at half maximum (FWHM), respectively.
The fitted profiles are almost comparable to the beam size, which means that the radial structure of the dust ring can be resolved only marginally by our observations.
Assuming a completely geometrically thin condition, the ring width along the minor axis should be reduced by a factor of $\cos i$ even if the ring width is only resolved marginally.
Thus, the similar values of the measured widths imply that the ring is not geometrically thin.
If we assume that the inferred emissions of the dust ring have a gaussian distribution, the beam-deconvolved widths along the major and minor axes, $w_\mathrm{maj}$ and $w_\mathrm{min}$, can be measured to be 0\farcs156 and 0\farcs150 in FWHM, corresponding to 24.6 and 23.7~au, respectively.
The detailed structure of the ring should be confirmed by further observations at higher resolution.

\subsection{CO(3--2) Emission}\label{sec:co32}
\begin{figure}
\begin{center}
	\includegraphics[width=\textwidth]{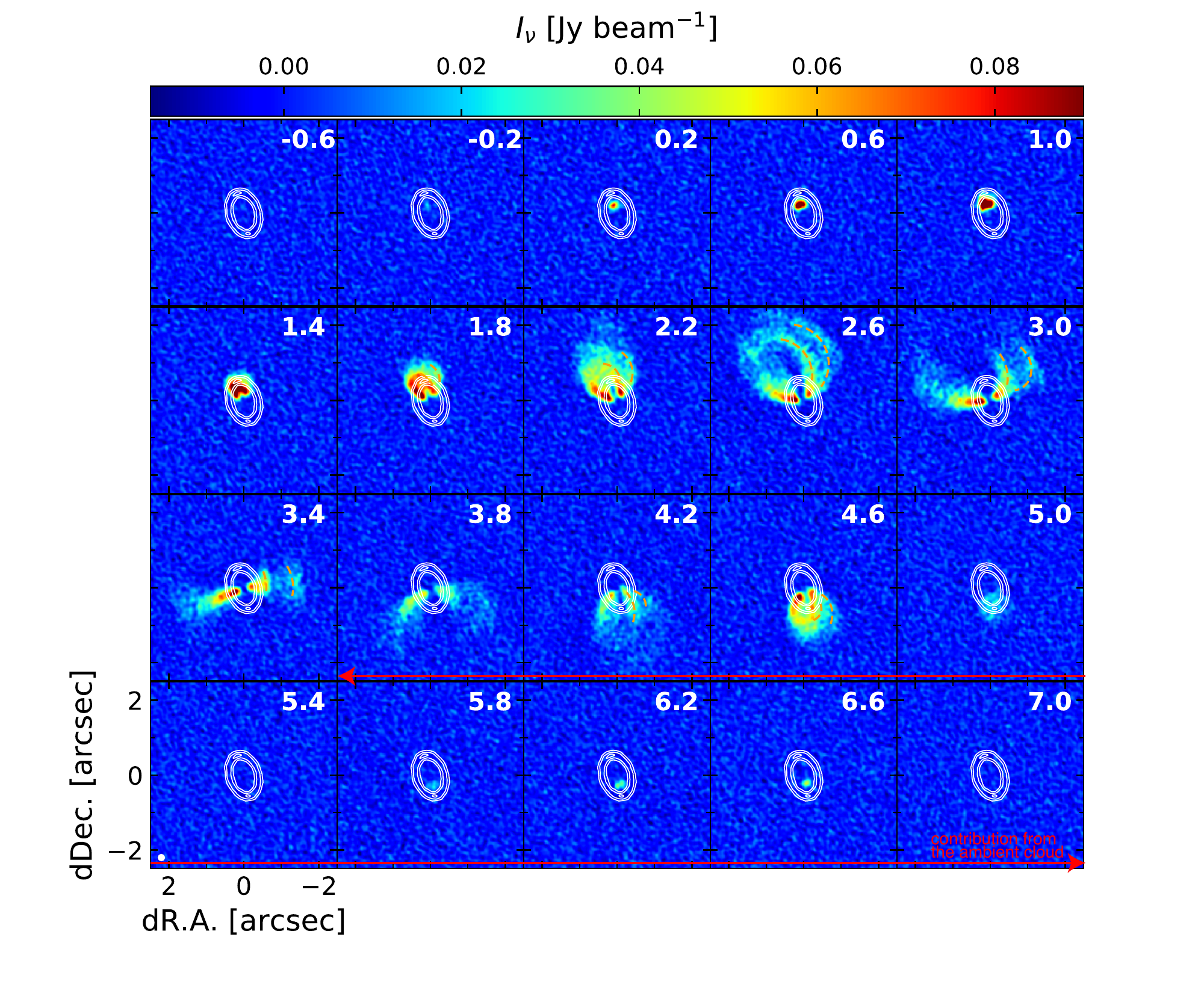}
	\caption{Channel maps of the CO(3--2) emission are shown in color. The white contour indicates the dust continuum emission with contour intervals of 10$\sigma$ (see Figure \ref{fig:b7cont}). The white circle at the bottom-left corner of each panel is the beam size. The central LSR velocity in km s$^{-1}$ is shown at the top-right corner in each panel. The twin-line patterns we mentioned in the text are traced by the dotted lines in orange. The velocity range where the emission is probably contaminated by the parent cloud is shown by the red arrow.}\label{fig:co_chmap}
\end{center}
\end{figure}

Figure \ref{fig:co_chmap} shows the velocity channel maps of the CO(3--2) line after channel binning to $\Delta v=0.4$ km s$^{-1}$ for visualization.
The CO(3--2) emission is detected above 3$\sigma$ in a velocity range from $v_\mathrm{LSR}=-0.2$ to 6.6~km s$^{-1}$, while the CO emission at $\sim$4--7 km s$^{-1}$ is likely to be contaminated by the filtered out effect due to the extended Lupus III cloud \citep{bib:tsukagoshi2014}.

An overall velocity gradient is seen along the major axis of the dust ring, implying that a rotating gas disk is inclined similarly to the dust ring.
Since the emissions at higher blueshifted and redshifted velocities with respect to the systemic velocity of the CO disk \citep[3.6 km s$^{-1}$;][]{bib:tsukagoshi2014} are located close to the star, the gas disk is most likely undergoing Keplerian rotation.
The CO gas emission clearly extends beyond the dust ring up to $2\arcsec$ in radius ($\sim$320~au), which is consistent with the previous CO images \citep{bib:tsukagoshi2014,bib:canovas2015}.
The CO gas emission also occurs in the inner hole of the dust ring, easily recognized by the higher velocity components.
Assuming that the gas disk is inclined by $49\degr$ from face-on and is undergoing Keplerian rotation, the maximum velocities of $-0.2$ and $6.6$~km s$^{-1}$ indicate that the gas disk extends to an inner radius of $\sim$8~au.
The inner radius we estimated is smaller than that of \citet{bib:canovas2015}, probably due to the improvement in sensitivity.
The total integrated intensity of the CO(3--2) emission is measured to be 7.6$\pm$0.1~Jy~km s$^{-1}$, which is consistent with the previous observation by \citet{bib:tsukagoshi2014} within 10\% uncertainty.

A complex twin-line velocity pattern tracing different CO-emitting regions can be recognized in each channel map from 1.8 to 4.6~km~s$^{-1}$, while the Keplerian rotation pattern is dominant.
At velocities 1.8--2.6 km s$^{-1}$, for example, there are two arm-like features extending to the north from the vicinity of the central star.
Such a twin-line pattern is reminiscent of the two-layered emissions showing the front and rear sides of a flared disk, as seen in the CO disks of HD~163296 \citep{bib:rosenfeld2013,bib:gregorio-mosalvo2013}, IM~Lup \citep{bib:pinte2017}, and HD~97048 \citep{bib:plas2017a}.
\replaced{In this case, the twin-line pattern observed only at the west side of the disk is consistent with the crescent-like feature of the scattered light at NIR due to the forward scattering \citep{bib:tsukagoshi2014}.
}{The twin-line pattern is found only at the west with respect to the central star, corresponding to the near side disk surface.
This fact agrees with the crescent-like feature of the scattered light at NIR in which the emission is bright at the west due to the forward scattering at the near side disk surface \citep{bib:tsukagoshi2014}.}
There is a contribution from the extended ambient cloud at $\sim$4--7 km s$^{-1}$ \citep{bib:tsukagoshi2014}, and thus the twin-line pattern is difficult to discern in this velocity range.

\begin{figure}
\begin{center}
	\includegraphics[width=\textwidth]{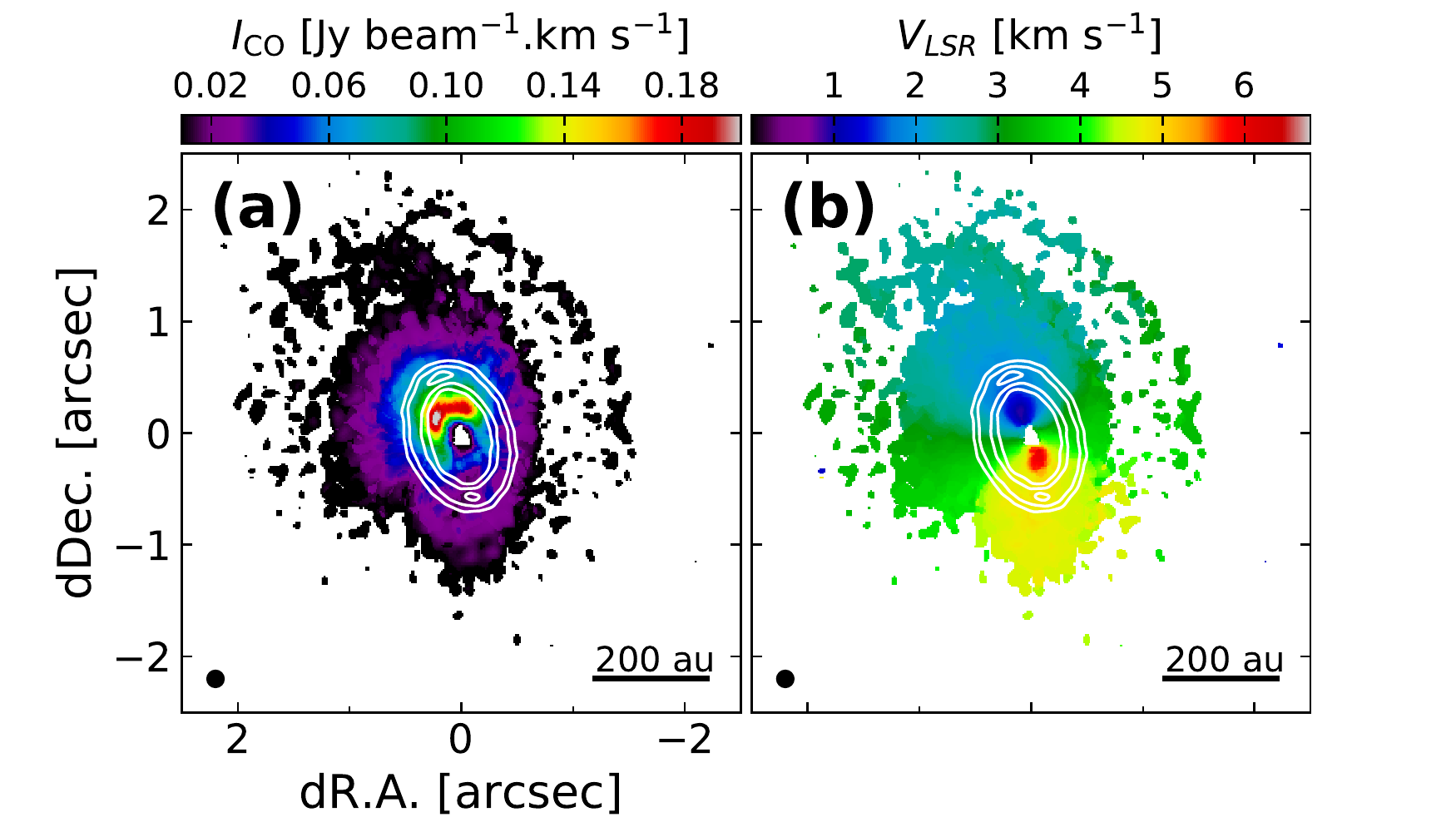}
	\caption{Moment maps of the CO(3--2) emission created using the $>5\sigma$ pixels. The panels (a) and (b) show the zeroth and first moments, respectively. The white contour indicates the dust continuum emission with contour intervals of 10$\sigma$ (see Figure \ref{fig:b7cont}). The filled circle at the bottom-left corner of each panel is the beam size.}\label{fig:co_moments}
\end{center}
\end{figure}

We made moment maps of the CO emission created by using the $>5\sigma$ data in the channel maps, as shown in Figure \ref{fig:co_moments}.
The integrated CO intensity peaks at the inner hole of the dust ring, as shown in the zeroth moment map (Figure \ref{fig:co_moments}a).
The asymmetry seen in the zeroth moment map is most likely due to the lack of the CO emission around 5.4 km s$^{-1}$.
The velocity gradient along the major axis of the dust ring is easily recognized in the first moment map (Figure \ref{fig:co_moments}b).
The position angle of the velocity gradient is estimated from the maximum and minimum velocity positions of the first moment map to be $30\degr$, which differs by 12$\degr$ from the position angle of the major axis of the dust ring (18$\degr$).
Note that the position angle is not seriously affected by the contribution from the ambient cloud because the emission at the highest velocity is so compact that the peak position is well determined. \par

\subsection{HCO$^{+}$(4--3) Emission}
\begin{figure}
\begin{center}
	\includegraphics[width=\textwidth]{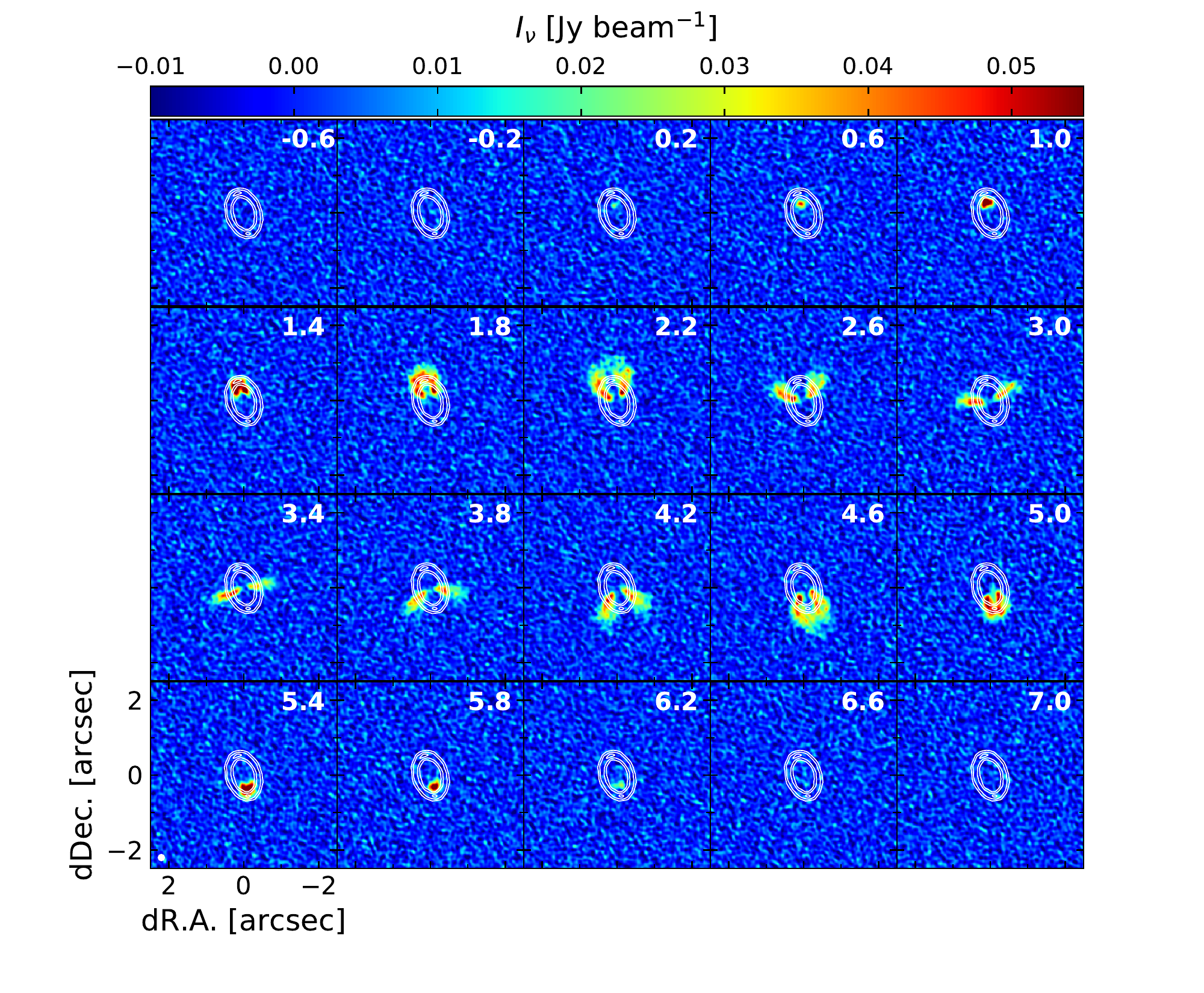}
	\caption{Same as Figure\ \ref{fig:co_chmap}, but for the HCO$^{+}$(4--3) emission.}\label{fig:hcop_chmap}
\end{center}
\end{figure}

\begin{figure}
\begin{center}
	\includegraphics[width=\textwidth]{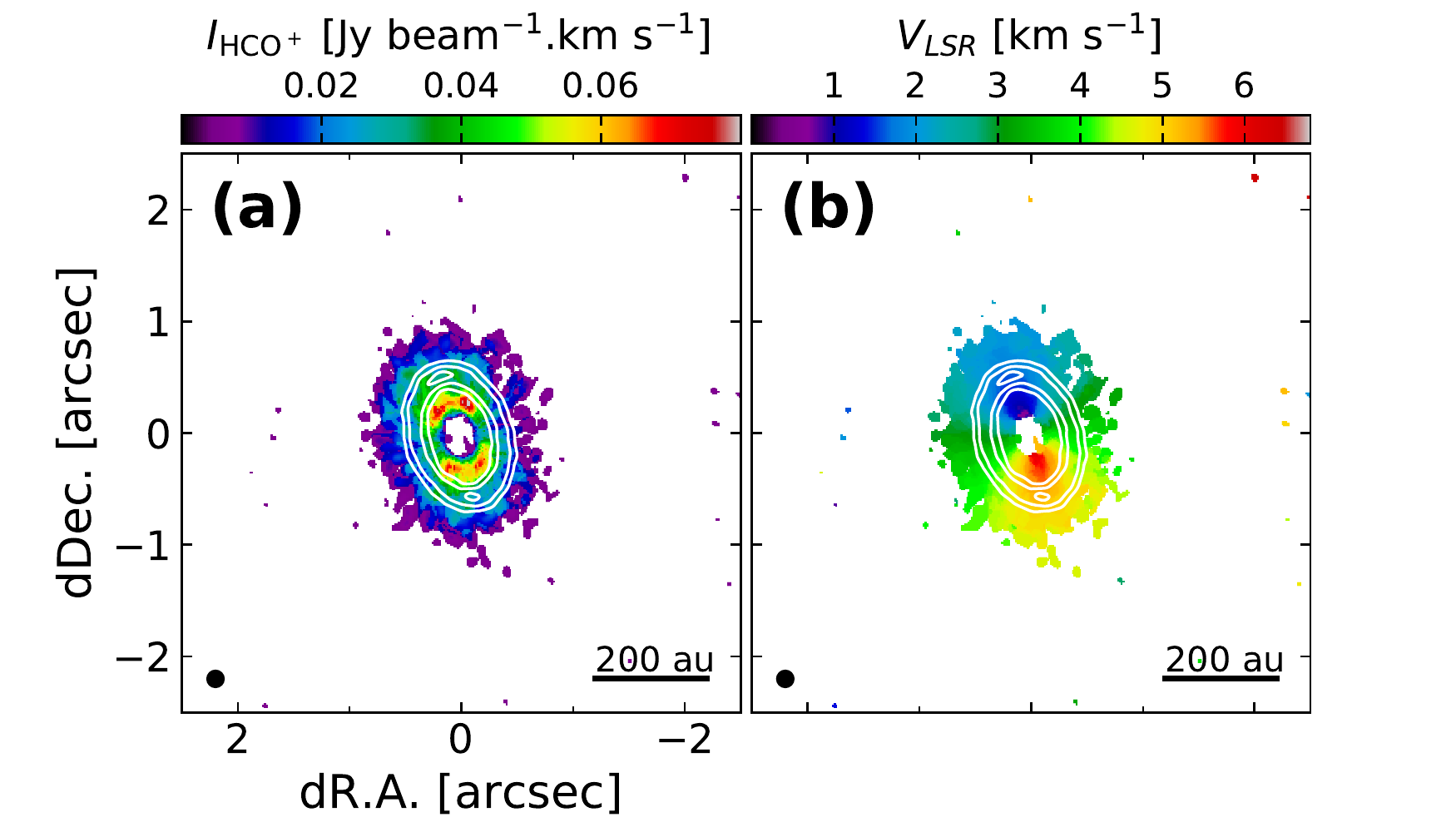}
	\caption{Same as Figure\ \ref{fig:co_moments}, but for the HCO$^{+}$(4--3) emission.}\label{fig:hcop_moments}
\end{center}
\end{figure}

Figure \ref{fig:hcop_chmap} shows the velocity channel maps of the HCO$^{+}$(4--3) line.
The $>4\sigma$ threshold is employed to create the moment maps because the signal-to-noise ratio is lower than that of the CO image, and the maps are shown in Figure \ref{fig:hcop_moments}.
The overall structure of HCO$^{+}$ is found to be similar to that of the CO emission.
However, no complex twin-line velocity pattern is found in the HCO$^{+}$ emission.
The HCO$^{+}$ emission is detected from $v_\mathrm{LSR}=$0.2 to 6.2~km~s$^{-1}$, and the overall velocity gradient is along the NE-SW direction.
The emission extends to the inner and outer parts with respect to the dust ring.
The total integrated intensity is measured to be 3.3$\pm$0.1~Jy~km s$^{-1}$.\par

In contrast to the CO emission, the zeroth moment map shows a symmetric distribution.
This is because the spatially filtered ambient cloud does not affect the HCO$^{+}$ emission whose critical density ($\sim6\times10^6$ cm$^{-3}$) is much higher than the density of the ambient gas \citep[$\sim$10$^{3}$ cm$^{-3}$;][]{bib:tachihara1996}.
The position angle of the velocity gradient is measured to be $19\degr$, which agrees well with the position angle of the dust ring and differs by 11$\degr$ from that of the CO emission.
This discrepancy of 11 deg probably reflects the difference in the height of each emitting region where the emissions become optically thick along the vertical direction of the disk.
Or, the gas disk might have a variation of angular momentum as a function of radius and is warped slightly.

\section{Discussion}\label{sec:discussion}
\subsection{Front and Rear Surfaces of the CO Disk}\label{sec:co_disk}
The twin-line pattern seen in the CO channel maps can be explained if the temperature of the disk midplane is so low that the CO line does not emit effectively.
In this case, only the disk surfaces at the front and rear are expected to be bright because the intensity of the optically thick CO emission depends on the temperature.
Such a cold midplane may be common in protoplanetary disks \citep[e.g.,][]{bib:chang1997}.
The dust grains in the Sz~91's disk are concentrated within a specific radius range, and the gas temperature is expected to be low behind the dust ring because it blocks the radiation from the central star.\par

To check whether the twin-line pattern can be reproduced by the combination of the dust ring and the flared gas disk, we performed radiative transfer calculations using RADMC-3D\footnote{RADMC-3D is an open code for radiative transfer calculations developed by Cornelis Dullemond. The code is available online at \url{http://www.ita.uni-heidelberg.de/~dullemond/software/radmc-3d/index.html}}.
The details of the disk model are described in \S\ref{sec:setup}, and the results of our calculation are summarized in \S\ref{sec:rtc}.
Note that this calculation is not aimed at a complete fitting to the emission.\par

\subsubsection{Model Setup}\label{sec:setup}
\begin{figure}
\begin{center}
	\includegraphics[width=0.6\textwidth]{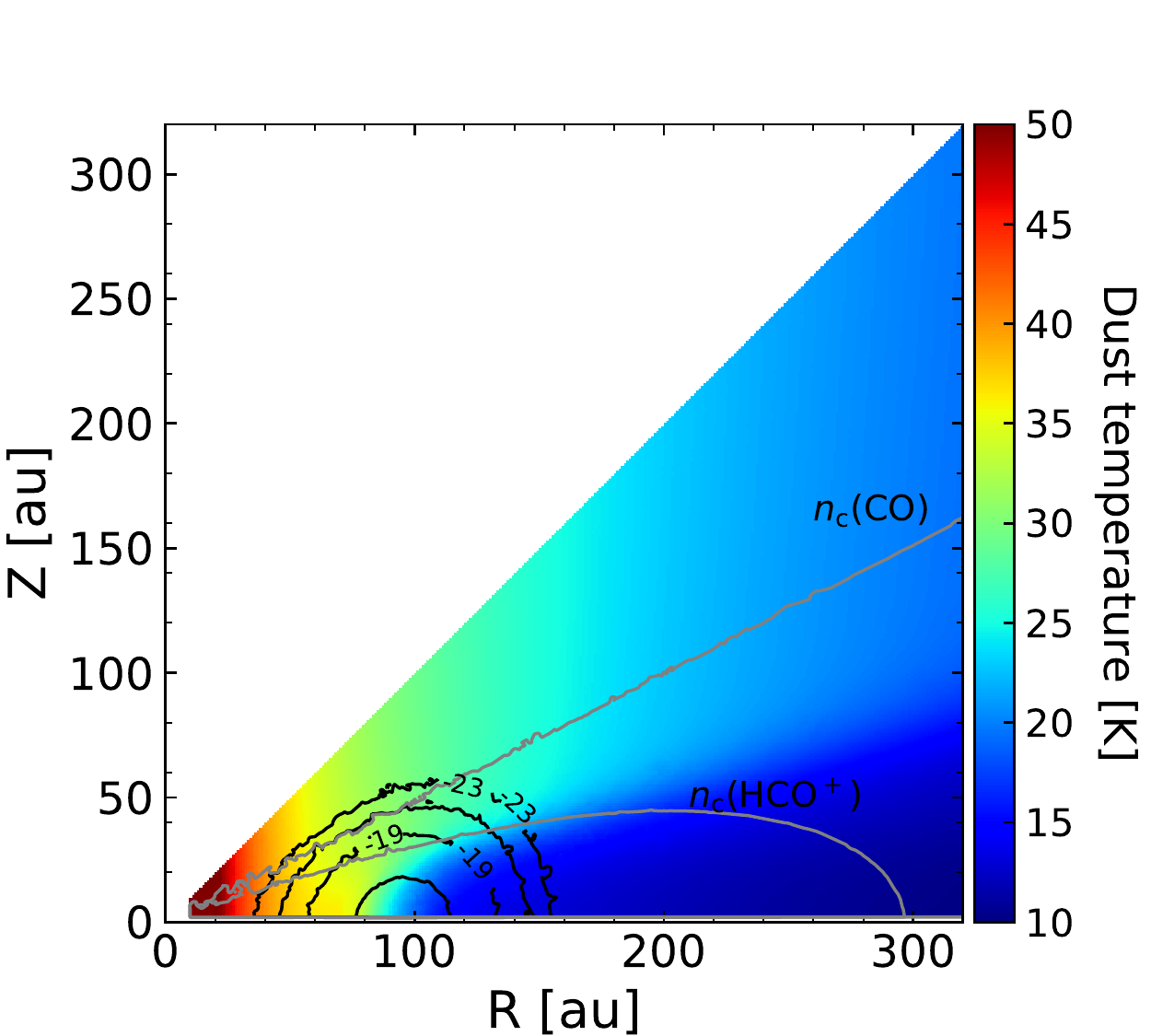}
	\caption{Temperature distribution obtained by the thermal Monte Carlo calculation, shown in color. The black contour indicates the density structure of the dust disk, which is the input model for the calculation. The logarithm value of the dust density (g~cm$^{-3}$) is labeled for each contour line. The gray lines indicate the boundary where the gas density exceeds the critical densities of CO and HCO$^{+}$.}\label{fig:model}
\end{center}
\end{figure}

We assume that the column density distribution of the dust ring is approximately a Gaussian function as described by
\begin{equation}
\Sigma_\mathrm{d}(r) = \Sigma_\mathrm{d,0} \exp \Biggl[ - \frac{(r-r_\mathrm{pk})^2}{2\sigma_\mathrm{r}^2} \Biggr] \ .
\end{equation}
Here $r_\mathrm{pk}$ is the peak position of the dust ring and $\sigma_\mathrm{r}$ is the standard deviation of the Gaussian.
We set $r_\mathrm{pk}$ and $\sigma_\mathrm{r}$ to 95~au and 10~au, respectively, based on our continuum image (see \S \ref{sec:result}).
The factor $\Sigma_\mathrm{d,0}$ is determined to be 1.84$\times$10$^{-2}$ g cm$^{-2}$ so that the total flux density is in reasonable agreement with the observed one.
The Gaussian distribution is also adopted along the vertical direction with the scale height parameter $h_\mathrm{d}$.
From the geometric consideration of the aspect ratio between the major and minor axes of the dust ring, $h_\mathrm{g}$ is supposed to be 9.8~au from the equation
\begin{equation}
h_\mathrm{g} = \frac{W_\mathrm{maj}}{2 \sin \theta} \sqrt{ \biggr( \frac{W_\mathrm{min}}{W_\mathrm{maj}} \biggr)^2 - \cos^2 \theta},
\end{equation}
where $W_\mathrm{maj}$ and $W_\mathrm{maj}$ are the FWHM widths along the major and minor axes.
Here, we assumed a constant $h_\mathrm{d}$ of $\sim$9.8~au, $\sim10$ \% of the $r_\mathrm{pk}$, throughout the dust ring.
The assumed $h_\mathrm{d}$ is $\sim$25 \% lower than the scale height of the gas disk in hydrostatic equilibrium, which can be estimated from $h_\mathrm{g}=c_\mathrm{s}/\Omega_\mathrm{K}$, where $c_\mathrm{s}$ is the sound speed and $\Omega_\mathrm{K}$ is the angular speed of the keplerian rotation.
If we adopt the temperature distribution in \citet{bib:tsukagoshi2014}, $h_\mathrm{g}$ is estimated to be 13.2~au at 95~au radius.
The $\sim$25\% dust settling is comparable to that determined directly for the edge-on Flying Saucer disk \citep{bib:guilloteau2016}.
The density of the dust ring $\rho_\mathrm{d}$ is thus expressed by
\begin{equation}
\rho_\mathrm{d}(r,z) =  \frac{\Sigma_\mathrm{d}(r)}{\sqrt{2 \pi} h_\mathrm{d}} \exp \Biggl(  -\frac{z^2}{2h_\mathrm{d}^2} \Biggr)\ ,
\end{equation}
and thus, the resultant density distribution shows a torus-like structure, as shown in Figure \ref{fig:model}.\par

\begin{figure}
\begin{center}
	\includegraphics[width=0.7\textwidth]{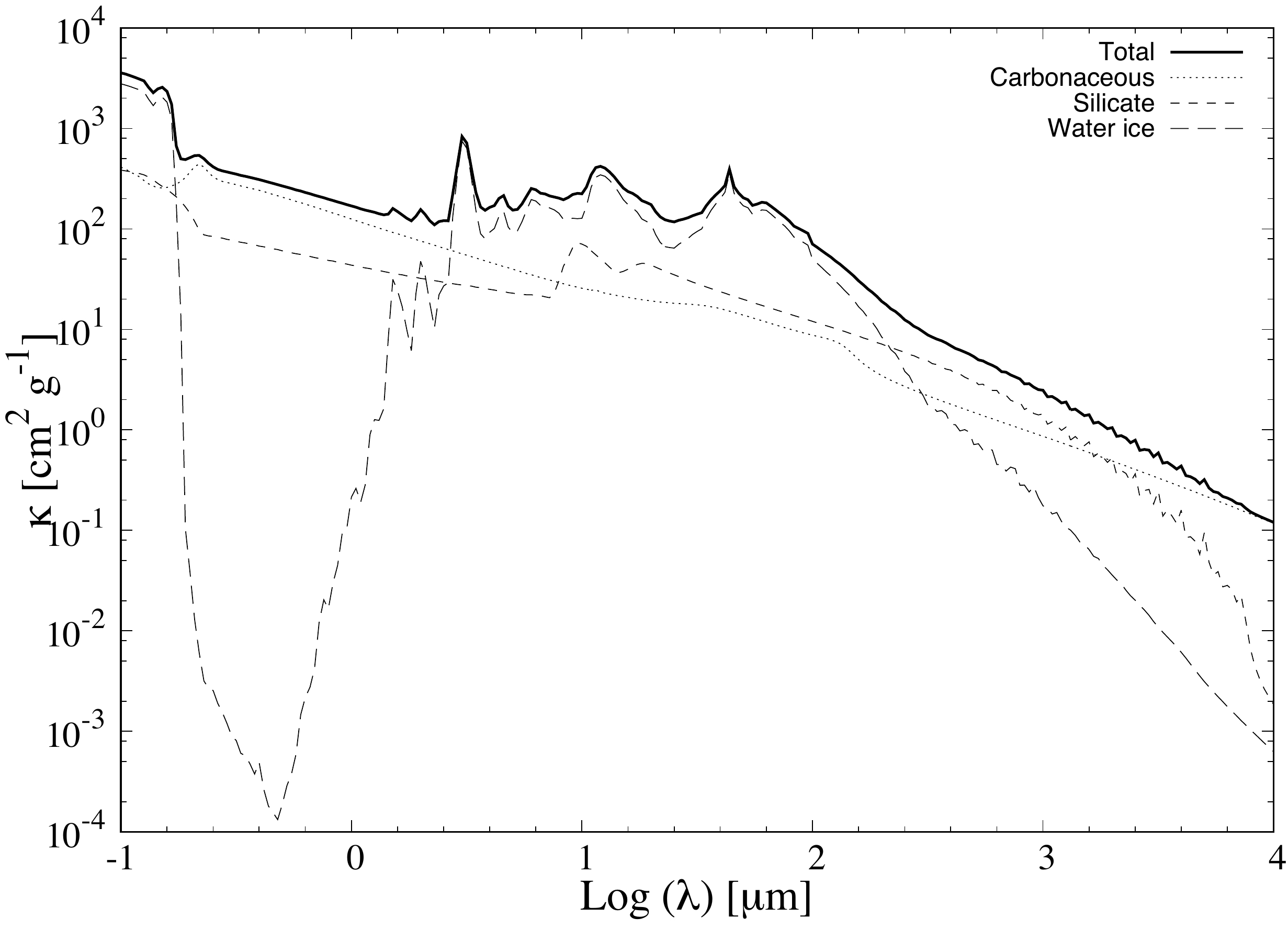}
	\caption{Dust opacities adopted in this study. The dotted, dashed, and long-dashed lines show the contribution from carbonaceous, silicate, and water ice grains, respectively. The thick solid line indicates the total dust opacity.}\label{fig:opac}
\end{center}
\end{figure}

The dust opacity used in our calculations follows that of \citet{bib:aikawa2006} and is shown in Figure \ref{fig:opac}.
The dust grains are assumed to be spherical with radius $a$, and are a mixture of silicate grains, carbonaceous grains, and water ice.
Their fractional abundances relative to the hydrogen mass are taken to be $\zeta_\mathrm{sil}=0.0043$, $\zeta_\mathrm{carbon}=0.0030$, and $\zeta_\mathrm{ice}=0.0094$, and their bulk densities are set to be $\rho_\mathrm{sil}=3.5$, $\rho_\mathrm{graphite}=2.24$, and $\rho_\mathrm{ice}=0.92$ g cm$^{-3}$, respectively \citep[for details see][]{bib:nomura2005}.
The dust grains are assumed to have a power-law size distribution given by
\begin{equation}
\frac{dn(a)}{da} \propto a^{-3.5}\ .
\end{equation}
The maximum grain radius $a_\mathrm{max}$ is set to be 1~mm in consideration of the dust growth relative to that in the interstellar medium.
The power-law slope of the opacity at longer wavelength is $\sim -1.2$.\par

The column density of the gas disk is assumed to have a truncated power-law form, as described by
\begin{equation}
\Sigma_\mathrm{g}(r) = \Sigma_\mathrm{g,100} \Biggl( \frac{r}{\mathrm{100~au}} \Biggr)^{-1.0}
\end{equation}
for 8$<r<$320 au.
Here, $\Sigma_\mathrm{g,100}$ is the column density at 100~au, which is chosen so that the total gas mass $M_\mathrm{g}$ maintains a gas-to-dust mass ratio of 100, i.e., $M_\mathrm{g} = 100 \times M_\mathrm{d}$.
This means that the local gas-to-dust mass ratio varies over the disk, and the minimum gas-to-dust mass ratio is $\sim6$ at the centroid of the dust ring.
Hydrostatic equilibrium along the vertical direction is assumed.
The scale height of the gas disk is assumed to be $h_\mathrm{g}=$13.2~au at $r=$95~au and has the same power-law index as a typical flared disk \citep[e.g.,][]{bib:hayashi1981}, which can be described as
\begin{equation}
h_\mathrm{g} = \mathrm{13.2} \Biggl( \frac{r}{\mathrm{95~au}} \Biggr)^{1.25}\ \mathrm{[au]}.
\end{equation}
Since the gas scale height of 13.2~au corresponds to $\sim24$~K at 95~au, it is consistent with the temperature profile derived by the thermal Monte Carlo calculation presented in Figure \ref{fig:model}.
Thus, the density of the gas disk is given by 
\begin{equation}
\rho_\mathrm{g}(r,z) = \frac{\Sigma_\mathrm{g,100}}{\sqrt{2 \pi}h_\mathrm{g}} \Biggl( \frac{r}{\mathrm{100au}} \Biggr)^{-1.0} \exp \Biggl(  -\frac{z^2}{2h_\mathrm{g}^2} \Biggr)\ .
\end{equation}
The molecular abundances of CO and HCO$^+$ with respect to H$_2$ are assumed to be constant over the disk, i.e., no condensation of the molecules onto the dust grains is assumed.
Those are $10^{-4}$ and $10^{-9}$ for CO and HCO$^+$, respectively \citep{bib:qi2011,bib:salter2011}.\par

The gas motion is assumed to be a Keplerian rotation whose velocity field is written by
\begin{equation}
V(r) = \Biggl( \frac{GM_\ast}{r} \Biggr)^{0.5}\ ,
\end{equation}
where $G$ is the gravitational constant.
The velocity broadening due to turbulence is assumed to be constant over the disk and is fixed to be 120 m s$^{-1}$, according to \citet{bib:canovas2015}.\par

The temperature distribution is obtained by the thermal Monte Carlo calculation using 10$^9$ photons.
The result of the calculation is shown in Figure\ \ref{fig:model}.
As predicted above, the lower temperature is actually achieved near the midplane behind the dust ring.
The obtained temperature distribution is likely valid because a similar trend is found in the model for another ring-like transitional disk using a different method \citep{bib:muto2015}.
In the radiative transfer calculation, we assume that the disk is in local thermodynamic equilibrium and the gas temperature is the same as the dust temperature ($T_\mathrm{gas}=T_\mathrm{dust}$).\par


\subsubsection{Radiative Transfer Calculation and Mock Observation}\label{sec:rtc}
The radiative transfer calculations were performed for a velocity range from $v_\mathrm{LSR}=-8$ to 8 km s$^{-1}$ with a 0.2 km s$^{-1}$ step, and the channel maps of CO and HCO$^{+}$, as well as the dust continuum emission at band 7, were made.
The contribution from the continuum emission in the channel maps of CO and HCO$^{+}$ was subtracted in the image domain.
After that, the modeled images for the continuum, CO(3--2) and HCO$^{+}$(4--3), were reprocessed to produce similar visibility samplings to the observed one using the {\it simobserve} task.
Then, the CLEANed images of the models were created by the {\it clean} task with the same imaging parameter as for the observation.\par

\subsubsection{Result Overview}
In summary, the results of our calculations indicate that the continuum emission can be reproduced by a ring-like distribution of dust grains with a maximum radius of 1~mm, and the twin-line pattern seen in the CO channel maps can be explained by a combination of the cold midplane caused by the dust ring and the flared gas disk without heavily depleted CO molecules.
Since some differences between the observed and simulated maps still remain, it is necessary to improve the disk model for a complete fitting to the observed data.
Dependence on the disk parameters and comparison with other disks are discussed in the following sections.

\begin{figure}
\begin{center}
	\includegraphics[width=\textwidth]{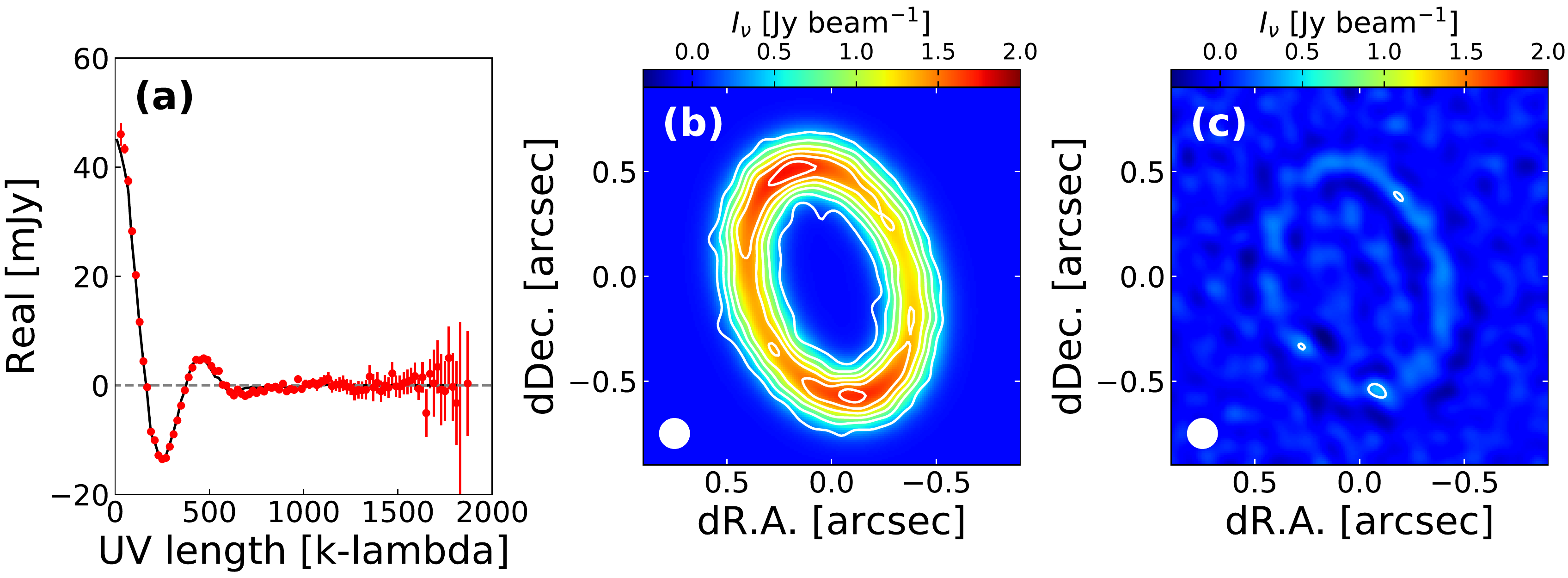}
	\caption{\added{Comparison of the modeled data by our radiative transfer calculation with the observed one. (a) Real part of the azimuthally averaged continuum visibility profile as a function of the deprojected baseline length. The visibility data is resampled by 20~k$\lambda$. The red circles represent the resampled visibility taken by our observations. The modeled visibility is shown by the solid line in black. (b) Modeled image reconstructed by CLEAN is shown in color. The color range is identical to the observed image shown in figure \ref{fig:b7cont}(a). The white contour represents the observed continuum emission with a contour interval of 5$\sigma$. (c) Subtracted image is shown in the same intensity range as in (b). The contour denotes 5$\sigma$.}}\label{fig:cont_sub}
\end{center}
\end{figure}

\replaced{
The band 7 continuum emission is well reconstructed in terms of the peak positions and the widths of the profiles, as shown in Figure\ \ref{fig:radplot}, while the intensity of the model image slightly lower than the observed one ($\sim$90\%).
The total dust mass is estimated to be $\sim$11 $M_\earth$, which is comparable to that of \citet{bib:canovas2015} as well as our estimation from the total flux density at 350 GHz in \S \ref{sec:result}.
}{ 
The band 7 continuum emission is well reconstructed by our fiducial model as shown in Figure \ref{fig:cont_sub}.
The visibility produced from the continuum image of the fiducial model is compared and subtracted from the observed one to obtain the residual map.
The CLEANed maps of the modeled and subtracted visibilities are created with the same imaging parameters as for the observed one.
It is clear in the residual map that most of the emission is subtracted and the residual at the ring is $\sim$15\% level at maximum.
This means that the asymmtric structure in the band 7 image is properly reproduced by the symmetric fiducial model.  
The peak position and the widths of the profiles also agree with the observed ones, as shown in Figure\ \ref{fig:radplot}, while the intensity of the model image slightly lower than the observed one ($\sim$90\%).
}

\begin{figure}
\begin{center}
	\includegraphics[width=\textwidth]{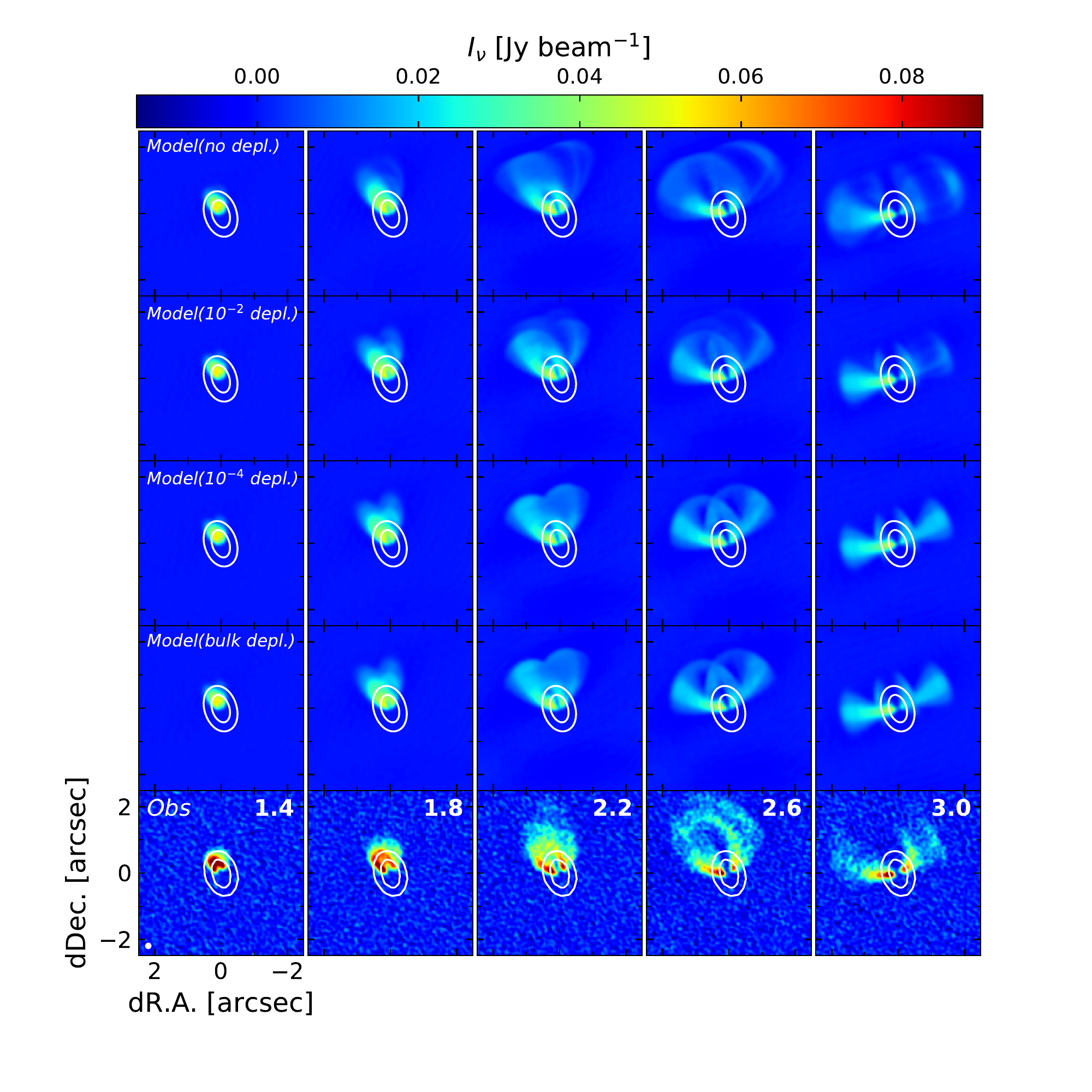}
	\caption{Comparison of the simulated CO emission from the modeled disk in the representative channels to the observed ones. The top row shows the simulated channel maps for the case where a constant molecular abundance is assumed throughout the disk. The following three rows also show the simulated channel maps, but for the case where molecules are depleted by a factor of 10$^{-2}$, 10$^{-4}$ and $\sim0$ in cold regions of $T< 20$ K. The bottom row shows the observed channel maps (Figure\ \ref{fig:co_chmap}). The color scales are identical in the all rows. The white contours represent the simulated continuum emission at band 7 for four rows from the top and the observed one for the bottom row. The contour intervals are 10$\sigma$, where 1$\sigma=$61.2 $\mu$Jy beam$^{-1}$ (figure \ref{fig:b7cont}).}\label{fig:co32model}
\end{center}
\end{figure}

In Figure \ref{fig:co32model}, we compare the CO channel maps obtained by our calculations to the observed ones on the same intensity scale.
Although there are some differences in the distribution and intensity, the model calculations seem to reproduce the observed twin-line pattern.
The expected CO intensity inside the dust ring is weaker than the observed one, indicating that a higher temperature than that expected in the model might be achieved in this region.

\begin{figure}
\begin{center}
	\includegraphics[width=\textwidth]{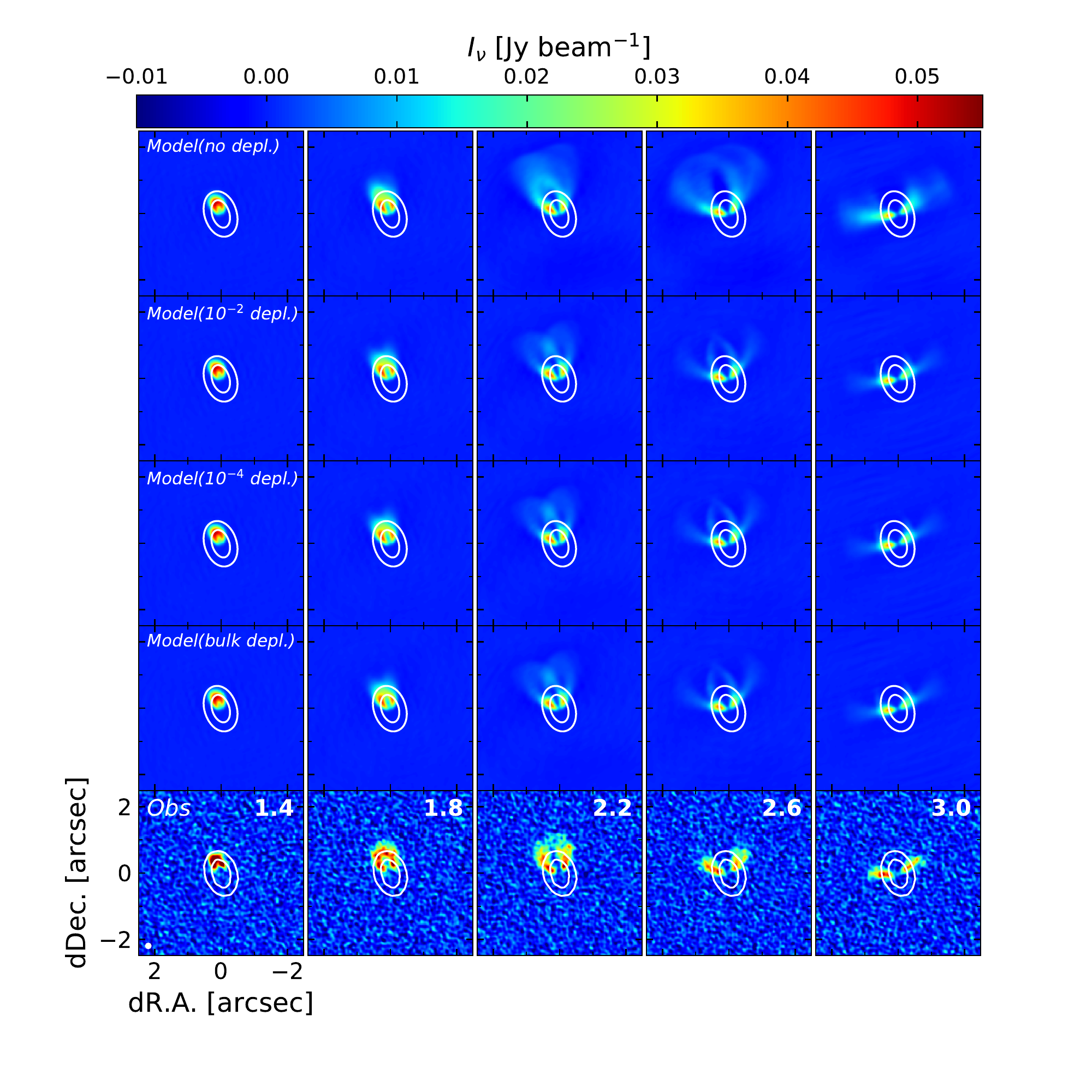}
	\caption{Same as Figure \ref{fig:co32model}, but for HCO$^{+}$.}\label{fig:hcop43model}
\end{center}
\end{figure}

Figure\ \ref{fig:hcop43model} shows the calculated channel maps of HCO$^{+}$, and their comparison with the observed ones.
The calculated HCO$^{+}$ emission shows the twin-line pattern as well and extends beyond the dust ring.
However, unlike the CO channels maps, this result is inconsistent with the observed ones; we can recognize the twin-line pattern in the simulated maps, while it does not appear in the observed one.
One of the possible reasons for this inconsistency is the uncertainty in the assumption of the total gas mass.
In our model, we simply assumed that the gas-to-dust mass ratio is fixed to 100 for the entire disk.
Thus, the H$_2$ gas density of the modeled disk is larger than the critical density of HCO$^{+}$ ($\sim6\times10^6$ cm$^{-3}$) at $r\lesssim300$~au in the midplane, while it is much larger than the critical density of CO ($\sim4\times10^4$ cm$^{-3}$) even in the outermost disk.
In other words, the observed weak intensities of HCO$^{+}$ outside the ring indicate that the actual H$_2$ gas density might be lower than that in the modeled disk.
If this is the case, the extent of the calculated HCO$^{+}$ emission should be smaller than that shown in Figure\ \ref{fig:hcop43model} because the boundary of the critical density is closer to the inner region.
For example, if the total H$_2$ mass is $\sim$10 \% of the assumed mass in the model, which is equivalent to a gas-to-dust mass ratio of 10 for the entire disk, the H$_2$ gas density is comparable to the critical density of the HCO$^{+}$ emission at $\sim95$~au, while still sufficient to excite the CO emission in the outermost disk.
This case seems to be more similar to the observed distributions of the CO and HCO$^{+}$ emissions, and thus a robust estimation of the gas-to-dust ratio is required to validate it.\par

The other possible reason for the mismatch in Figure \ref{fig:hcop43model} is that the HCO$^{+}$ molecules are more concentrated at smaller radii and near the midplane, as predicted by theoretical calculations \citep[e.g.,][]{bib:aikawa1999}.
If this is the case, the HCO$^{+}$ emission would be concentrated close to the central star and no twin-line pattern would exist.\par

\subsubsection{Parameter Dependence}
In our calculations, we assumed constant molecular abundances for the entire disk, i.e., no depletion occurs in the disk.
To check the effect of the depletion on the dust grains, we also ran the radiative transfer calculations under the assumption that the CO molecules are depleted by a factor of 10$^{-2}$, 10$^{-4}$ and $\sim0$ in the cold region where the disk temperature is below the sublimation temperature of CO ($<$20~K).
Note that we assume an extremely depleted disk since the CO depletion factor of a protoplanetary disk has been measured to be $\sim100$ \citep{bib:favre2013}.
The 10$^{-4}$ case of our calculation is similar to the partial depletion case of \citet{bib:pinte2017}.
The same degree of depletion for the HCO$^{+}$ molecules was adopted because the two species seem to coexist.\par

The resulting images are also shown in the middle rows of Figures \ref{fig:co32model} and \ref{fig:hcop43model}.
In the heavily depleted cases (depletion of 10$^{-4}$ and $\sim$0), the simulated CO images show that the gap between the twin lines tends to be diluted, and the twin-line pattern becomes symmetric with respect to the major axis of the gas disk.
This trend is easy to identified in the channel maps at 3.0~km~s${-1}$, which corresponds to the minor axis of the disk.
It is probable that self-absorption due to the cold CO gas near the midplane is insufficient due to the depletion to make the twin-line pattern asymmetric.
In other words, the CO emission from the warm surface region ($T>20$ K) behind the midplane can be seen directly, resulting in the symmetric twin-line patterns, while the emission from the rear side is heavily absorbed by the cold CO gas ($T<20$ K) near the disk midplane in the case of no depletion.
On the other hand, in the weak depletion case (depletion of 10$^{-2}$) and the fiducial model (no depletion), the twin-line pattern can be recognized due to the relatively clear gap between the twin lines, which seems to be in agreement with the observed one.
This trend is consistent with the case of the IM~Lup disk in which the CO molecules should be partially depleted to reproduce the observed morphology rather than being fully depleted \citep{bib:pinte2017}.\par

In contrast to the CO emission, the twin-line pattern in the simulated HCO$^{+}$ images seems to be symmetric for all the depletion cases while the asymmetric pattern can be found in the fiducial model.
Since a region whose density is comparable to the critical density of HCO$^{+}$ exists near the 20~K boundary, as shown in Figure\ \ref{fig:model}, it is expected that a large amount of the HCO$^{+}$ molecules are depleted and the HCO$^{+}$ emission becomes optically thin.
In this situation, the HCO$^{+}$ emission reflects the column density only along the nondepleted region.
Therefore, only the rims of the disk emission are brightened and the clear gap between the twin-line pattern appears.\par

\begin{figure}[htb]
\begin{center}
	\includegraphics[width=0.8\textwidth]{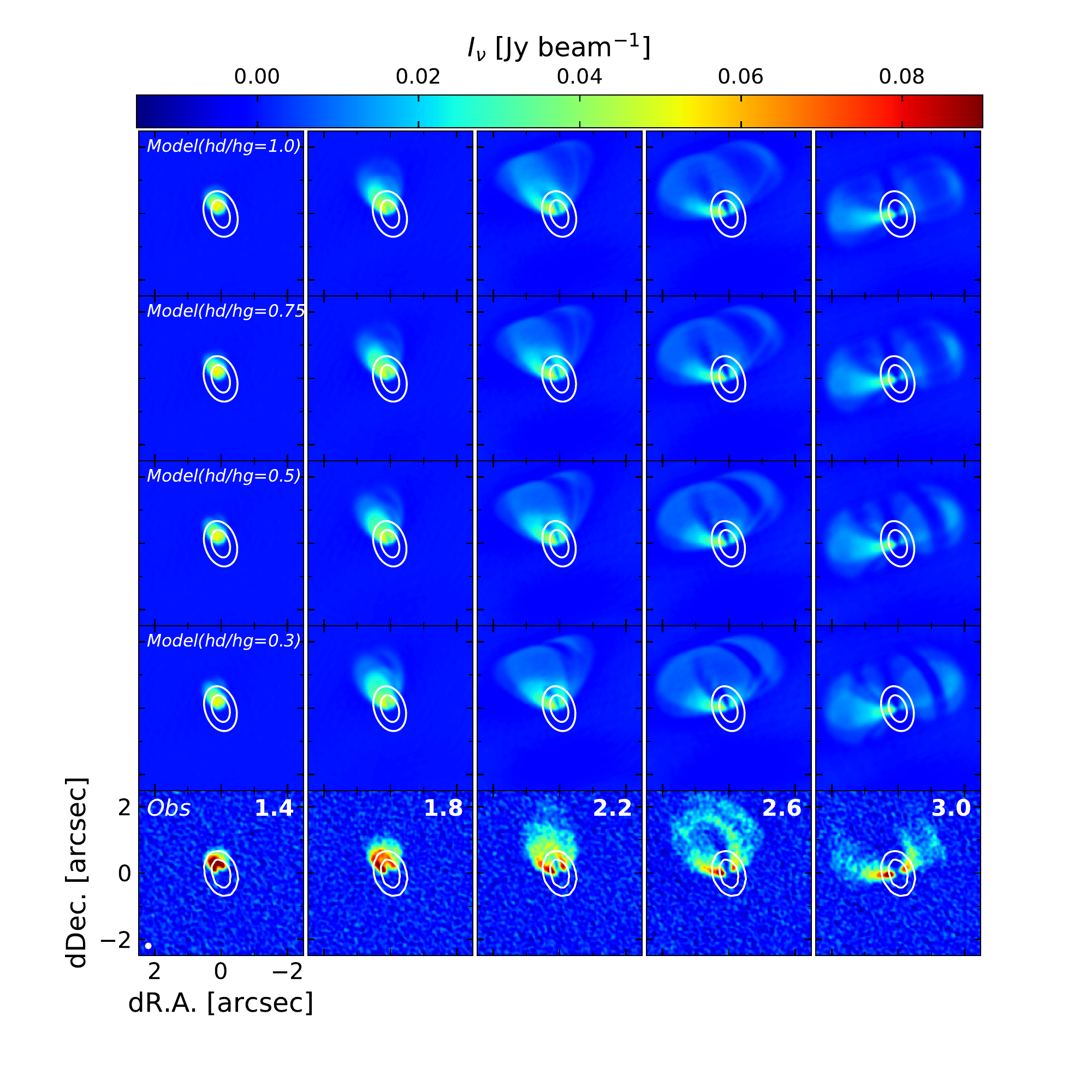}
	\caption{\replaced{Simulated CO emission in the representative channels are shown from the top to the 5th row when the dust scale height is varied to be $h_\mathrm{d}/h_\mathrm{g}=$1.0, 0.75 (fiducial model), 0.5, 0.3, and 0.1, respectively.}{Simulated CO emission in the representative channels are shown from the top to the 4th row when the dust scale height is varied to be $h_\mathrm{d}/h_\mathrm{g}=$1.0, 0.75 (fiducial model), 0.5, and 0.3, respectively.} The bottom row shows the observed channel maps (Figure\ \ref{fig:co_chmap}). The color scales are identical in the all the rows. The white contours represent the simulated continuum emission at band 7 for five rows from the top and the observed one for the bottom row. The contour denotes 10$\sigma$, where 1$\sigma=$61.2 $\mu$Jy beam$^{-1}$.}\label{fig:co32model_hd}
\end{center}
\end{figure}

\begin{figure}
\begin{center}
	\includegraphics[width=\textwidth]{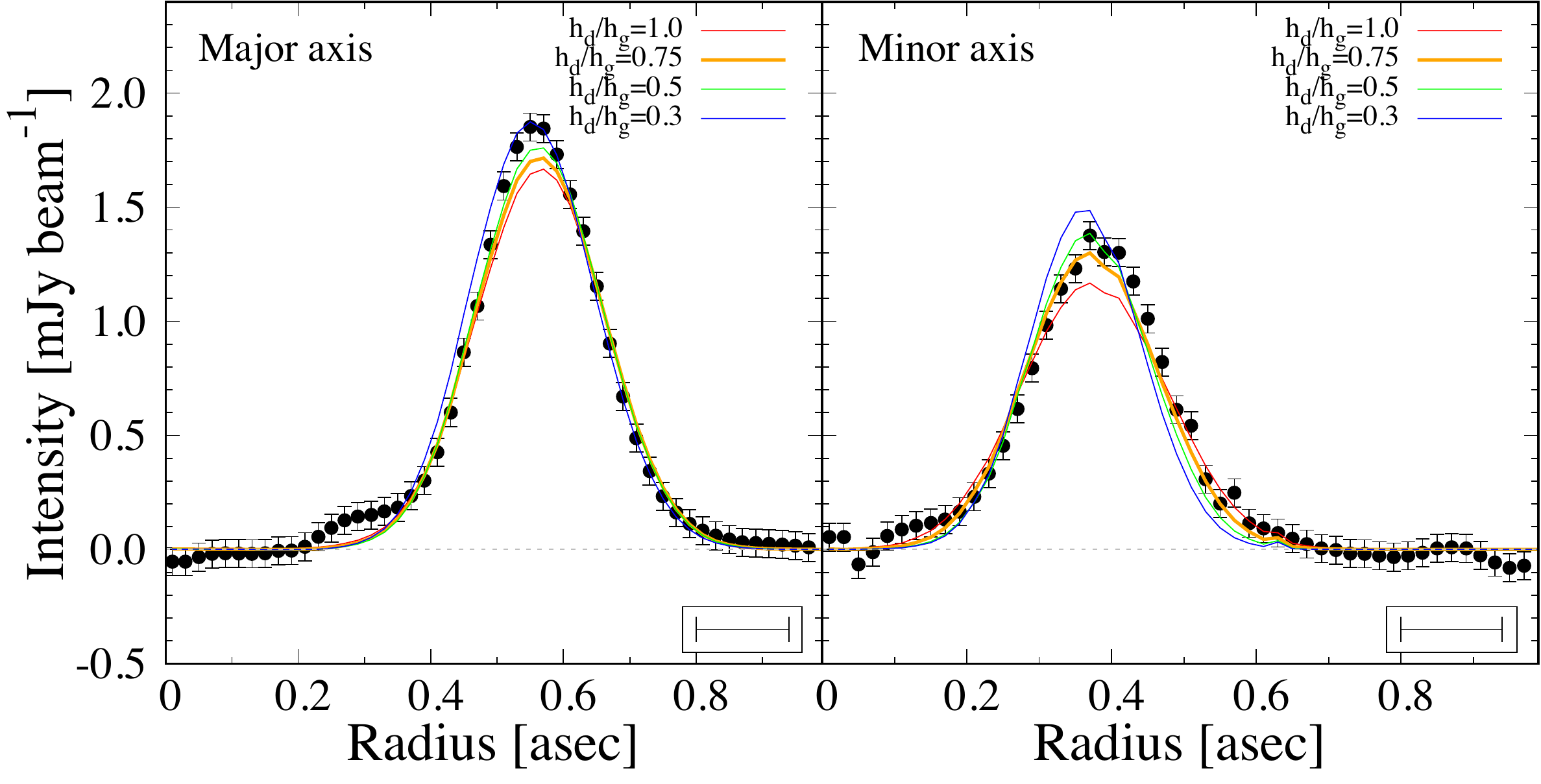}
	\caption{\replaced{Radial profiles extracted from the simulated band 7 continuum images by the same way as for the observed ones (Figure \ref{fig:radplot}). The profiles for the cases of $h_\mathrm{d}/h_\mathrm{g}=$1.0, 0.75, 0.5, 0.3, and 0.1 are shown in red, orange, green, blue and purple, respectively.}{Radial profiles extracted from the simulated band 7 continuum images by the same way as for the observed ones (Figure \ref{fig:radplot}). The profiles for the cases of $h_\mathrm{d}/h_\mathrm{g}=$1.0, 0.75, 0.5, and 0.3 are shown in red, orange, green, and blue, respectively.} The black filled circles show the observed ones. The bar at the bottom-right corner shows the FWHM of the synthesized beam ($0\farcs14$).}\label{fig:cont_model_hd}
\end{center}
\end{figure}

If different values of $h_\mathrm{d}$ are adopted, the separation between the twin-line pattern is changed.
This is because the shadowed region behind the dust ring against the stellar irradiation is varied.
\replaced{To check this, we run the radiative transfer calculations for the cases of $h_\mathrm{d}=$13.2, 6.6, 4.0, and 0.98~au in addition to the fiducial case (9.8~au), corresponding to $h_\mathrm{d}/h_\mathrm{g}=$1, 0.5, 0.3, and 0.1, respectively.
}{To check this, we run the radiative transfer calculations for the cases of $h_\mathrm{d}=$13.2, 6.6, and 4.0 in addition to the fiducial case (9.8~au), corresponding to $h_\mathrm{d}/h_\mathrm{g}=$1, 0.5, and 0.3, respectively.}
In each case, the $\Sigma_\mathrm{d,0}$ is set to have a comparable total flux density to the observed one within 2\%.
The results of the calculations show that the thinner the dust ring is, the narrower the separation becomes as shown in Figure \ref{fig:co32model_hd}.
The separations in the lower $h_\mathrm{d}$ cases are more similar to the observed one than that in the fiducial model, but there still remains uncertainty associated with the gas density distribution.
In contrast to the molecular line emissions, there is some inconsistency in the band 7 continuum data for the lower $h_\mathrm{d}$ cases.
\replaced{The radial profiles extracted from the simulated band 7 continuum image are consistent with the observed ones except for the lowest $h_\mathrm{d}$ case, as shown in Figure \ref{fig:cont_model_hd}.
}{
The radial profiles extracted from the simulated band 7 continuum image are consistent with the observed ones, as shown in Figure \ref{fig:cont_model_hd}.}
\deleted{For the lowest $h_\mathrm{d}$ case, which corresponds to the degree of dust settling for the \mbox{HL~Tau} case \citep{bib:pinte2016}, the dust mass is required to be $\sim2\times10^{-3}$~$M_\sun$ to reproduce the observed flux density, which is unrealistically large enough to be gravitationally unstable if the gas-to-dust ratio is assumed to be 100.
This is because the emission is optically thick ($\sim3$) under the lowest $h_\mathrm{d}$ condition, and thus the dust density must be considerably high to reach an agreement on the flux density.}
We also find that the observed aspect ratio of the FWHMs of the radial profiles (0.977$\pm$0.007), which is likely related to the dust scale height, agrees well with that derived from the fiducial model.
\replaced{From the model images of the continuum emission, we obtain 1.034, 0.946, 0.878, 0.837, and 0.848 for $h_\mathrm{d}/h_\mathrm{g}=$1.0, 0.75(fiducial), 0.5, 0.3, and 0.1, respectively, with a typical error  of 0.003.
}{From the model images of the continuum emission, we obtain 1.034, 0.946, 0.878, and 0.837 for $h_\mathrm{d}/h_\mathrm{g}=$1.0, 0.75(fiducial), 0.5, and 0.3, respectively, with a typical error  of 0.003.}
Note that the estimates are tentative because the width of the dust ring is comparable to the beamsize.\par

Although the turbulent velocity width results in the smearing of the emission in the channel maps, the reproduced twin-line pattern is not significantly affected by it.
To check how the turbulent velocity width affects the distribution, we made CO channel maps with turbulent velocity widths of 40, 80 and 240 m~s$^{-1}$.
We found that the twin-line pattern can be recognized even in the most turbulent case (240 m~s$^{-1}$), in which the channel map is much smeared by the turbulent velocity width.
In the less turbulent cases (40 and 80 m~s$^{-1}$), we also recognize the twin-line pattern and obtained similar channel maps to those obtained for the original case ($v_\mathrm{turb}=120$ m~s$^{-1}$).
This is because the emissions are less affected by turbulence than the smearing by the synthesized beam.\par

Finally, we check whether the inner radius $R_\mathrm{in}$ of the gas density distribution is valid or not.
In the fiducial model, we assume $R_\mathrm{in}=$8~au on the basis of the velocity difference between the blue- and red-shifted emissions (see \S\ref{sec:result}).
We also calculate the two cases of $R_\mathrm{in}=$~16 and 24~au and compare them to the observed data, focusing on the blue-shifted emission near the highest velocity ($-0.2$~km~s$^{-1}$).
As result, the CO emission above 3$\sigma$ is reproduced at the $-0.2$~km~s$^{-1}$ channel only for the $R_\mathrm{in}=$8~au case, while no emission is found for the $R_\mathrm{in}=$16 and 24~au models.
This result indicates that $R_\mathrm{in}$ is likely to be at least $<$16~au, which is smaller than the estimate in \citet{bib:canovas2015}.
The smaller $R_\mathrm{in}$ is probably due to the improvement in sensitivity as described in \S\ref{sec:result}.

\subsubsection{Comparison with Other Disks}
The two-layered asymmetric distribution that indicates the emission from the front and rear surfaces of the disk has been pointed out for some sources \citep{bib:rosenfeld2013,bib:gregorio-mosalvo2013,bib:pinte2017,bib:plas2017a,bib:plas2017b}.
A vertical temperature gradient is probably responsible for the distribution because in this situation the gas at the disk midplane is too cold not to emit effectively and most of the emission is from the warm surface layer.
The vertical temperature gradient is also likely the case for the Sz~91 disk as shown in figure \ref{fig:model}.
However, there is an inconsistency with the Sz~91 case (see the channel map at 3.0 km s$^{-1}$ in figure \ref{fig:co32model}).
If the disk emits only at the warm surface layer, the emission from both the front and rear surfaces would appear at the same position with the same radial velocity and no twin-line pattern would be observed along the minor axis.
One way to reproduce the twin-line pattern along the minor axis is that significant self-absorption due to the cold gas near the midplane should be taken into account.
In fact, our calculations demonstrate that the gap between the twin-line pattern is faded out for the heavily depleted cases, in which a large amount of cold CO gas is depleted.
We thus conclude that the vertical temperature gradient is crucial for the two-layered emission, and the remnant cold gas near the midplane also has a significant contribution to form the twin-line pattern.

\subsection{No Detection of the Hot Component}\label{sec:hot_component}
We did not find any counterparts of the hot component in the inner hole of the disk, which has been introduced to reproduce the excess emission in mid-IR in the SED \citep{bib:tsukagoshi2014}.
Adopting that the typical temperature in the cavity is 180~K and the dust mass opacity at the observed frequency is 2.8~cm$^2$~g$^{-1}$ \citep[][see also Figure\ \ref{fig:opac}]{bib:aikawa2006}, the 3$\sigma$ upper limit of the image (1$\sigma$=61.2 $\mu$Jy beam$^{-1}$) is equivalent to a dust mass of $\sim$4~$\times 10^{-3}~M_\earth$, while in \citet{bib:tsukagoshi2014}, the dust mass of the hot component has been predicted to be at least $>1.4\times10^{-3}~M_\earth$ adopting the same dust mass opacity and distance as in this study.
\added{Note that the dust mass opacity coefficient we assumed here is for a maximum grain size of 1~mm.
Smaller grains might be dominant at the inner disk because larger grains could be trapped efficiently in a pressure bump at the inner edge of the disk.
If a smaller dust mass opacity coefficient is adopted, which is expected for a smaller maximum grain size, the dust mass should be larger.}\par

The temperature corresponds to $\sim2$~au assuming the power-law temperature dependence in \citet{bib:tsukagoshi2014}, meaning that the hot component should be much smaller than the beamsize of this study.
Moreover, this radius is also smaller than the inner radius of the gas disk ($\sim$8~au), which was roughly estimated from the velocity dispersion in \S\ref{sec:co32}.
These results imply that the hot component with a mass of $\sim$1.4--4~$\times10^{-3}~M_\earth$ is located in the gas-depleted region.
Higher spatial resolution observations at mid-IR, as well as at submillimeter wavelengths, will provide new insights about the innermost structure.\par

\section{Summary}\label{sec:summary}
We present the results of high-spatial resolution and high-sensitivity aperture synthesis observations with ALMA of a transitional disk object in Lupus, \object{Sz~91}.
The dust continuum emission at band 7, and the molecular line emission of CO(3--2) and HCO$^{+}$(4--3) were observed, and these emissions were successfully imaged with a $0\farcs14$ resolution, equivalent to 22~au.
The main findings and conclusions are summarized as follows.

\begin{itemize}
\item{
We find that the dust continuum emission appears to be a symmetric ring, as previously reported \citep{bib:canovas2016}.
From a Gaussian fit to the radial profile, we find that the continuum emission peaks at a radius $\sim95$~au, and the beam-deconvolved width of the ring is measured to be 25~au.
The position and inclination angles of the disk are derived from the visibility fitting to be $18\fdg1\pm0\fdg2$ and $49\fdg7\pm0\fdg2$, respectively.
No significant emission is found in the inner hole even though the sensitivity is $\sim$10 times better than that in the previous ALMA image.
}
\item{
The detailed distribution of the molecular line emissions in CO and HCO$^{+}$ are revealed.
The emissions are detected at $v_\mathrm{LSR}=$ $-$0.2--6.6 km s$^{-1}$ and 0.2--6.2 km s$^{-1}$ for CO and HCO$^{+}$, respectively.
The velocity gradients are along the major axis of the dust emission, and the overall velocity trends agree well with the Keplerian rotation.
The CO emission extends to $\sim400$~au beyond the dust ring, as reported in the previous studies \citep{bib:tsukagoshi2014,bib:canovas2015,bib:canovas2016}.
The HCO$^{+}$ emission also extends beyond the dust disk, but appears to have a smaller size than that of CO.
We find in the channel maps of the CO emission that there are twin-line patterns showing the front and rear sides of a flared disk, while there are none in the HCO$^{+}$ emission.
}
\item{
To examine whether or not the twin-line patterns seen in the CO emission can be naturally explained by the combination of the dust ring and the flared gas disk, we performed radiative transfer calculations using RADMC-3D.
The disk model we constructed is based on our observational results, and the temperature distribution was obtained from the thermal Monte Carlo calculation.
Under the assumption of local thermodynamic equilibrium, the radiative transfer calculations for the continuum emission at band 7 and the molecular lines were done.
}
\item{
The thermal structure of the disk shows that there is a temperature gradient along the vertical direction beyond the dust ring.
This is because the dust ring blocks the stellar radiation.
We find that the observed twin-line pattern can be reproduced well by our disk model including the dust ring.
}
\item{
Our calculations indicate that heavy depletion of the CO molecules beyond the dust ring is unlikely.
This is because the CO emission from the warm surface layer on the disk rear side must be absorbed effectively at the cold midplane to achieve a reasonable agreement with the observed twin-line pattern.
No or weak depletion of the CO molecules is also supported by the fact that only a small amount of dust grains exists outside of the dust ring; it might be insufficient for the CO molecules to freeze-out onto dust grains effectively.
}
\item{
We could not find the counterpart of the hot component introduced by \citet{bib:tsukagoshi2014} for reproducing the excess emission at mid-IR.
The upper limit of our observations indicates that the total dust mass would be in the range of 1.4--4$\times10^{-3}$ M$_\earth$ if the hot component is an optically thick ring.
Since the expected radius of the hot component of $\sim$2~au is smaller than the inner radius of the gas disk derived in this study ($\sim8$~au), the hot component is expected to be located in the gas-depleted region in the innermost disk.
Further observations are needed to address this structure.
}
\end{itemize}

\acknowledgments
We appreciate the referee for the constructive comments that have helped to improve this manuscript.
We also thank Hideko Nomura for providing us with the dust opacity table used in this paper.
This paper makes use of the following ALMA data: ADS/JAO.ALMA\#2012.1.00761.S.
ALMA is a partnership of ESO (representing its member states), NSF (USA) and NINS (Japan), together with NRC (Canada), NSC and ASIAA (Taiwan), and KASI (Republic of Korea), in cooperation with the Republic of Chile.
The Joint ALMA Observatory is operated by ESO, AUI/NRAO and NAOJ.
A part of the data analysis was carried out on the common-use data analysis computer system at the Astronomy Data Center of NAOJ.
This work is partially supported by JSPS KAKENHI grant numbers 24103504 and 17K14244.



\facilities{Atacama Large Millimeter/submilliemeter Array}


\listofchanges


\begin{thebibliography}{}


\bibitem[Aikawa \& Herbst(1999)]{bib:aikawa1999}
Aikawa, Y., \& Herbst, E.\ 1999, \aap, 351, 233

\bibitem[Aikawa \& Nomura(2006)]{bib:aikawa2006}
Aikawa, Y., \& Nomura, H.\ 2006, \apj, 642, 1152

\bibitem[Alcal{\'a} et al.(2017)]{bib:alcala2017}
Alcal{\'a}, J.~M., Manara, C.~F., Natta, A., et al.\ 2017, \aap, 600, A20

\bibitem[Andrews \& Williams(2005)]{bib:andrews2005}
Andrews, S.~M., \& Williams, J.~P.\ 2005, \apj, 631, 1134 

\bibitem[Andrews et al.(2011)]{bib:andrews2011}
Andrews, S.~M., Wilner, D.~J., Espaillat, C., et al.\ 2011, \apj, 732, 42

\bibitem[Andrews et al.(2016)]{bib:andrews2016}
Andrews, S.~M., Wilner, D.~J., Zhu, Z., et al.\ 2016, \apjl, 820, L40

\bibitem[Bailer-Jones et al.(2018)]{bib:bailer-jones2018}
Bailer-Jones, C.~A.~L., Rybizki, J., Fouesneau, M., Mantelet, G., \& Andrae, R.\ 2018, arXiv:1804.10121

\bibitem[Brown et al.(2009)]{bib:brown2009}
Brown, J.~M., Blake, G.~A., Qi, C., et al.\ 2009, \apj, 704, 496


\bibitem[Casassus et al.(2013)]{bib:casassus2013}
Casassus, S., van der Plas, G., M, S.~P., et al.\ 2013, \nat, 493, 191

\bibitem[Chiang \& Goldreich(1997)]{bib:chang1997}
Chiang, E.~I., \& Goldreich, P.\ 1997, \apj, 490, 368


\bibitem[Canovas et al.(2015)]{bib:canovas2015}
Canovas, H., Schreiber, M.~R., C{\'a}ceres, C., et al.\ 2015, \apj, 805, 21

\bibitem[Canovas et al.(2016)]{bib:canovas2016}
Canovas, H., Caceres, C., Schreiber, M.~R., et al.\ 2016, \mnras, 458, L29


\bibitem[Cieza et al.(2010)]{bib:cieza2010}
Cieza, L.~A., Schreiber, M.~R., Romero, G.~A., et al.\ 2010, \apj, 712, 925

\bibitem[Cieza et al.(2012)]{bib:cieza2012}
Cieza, L.~A., Schreiber, M.~R., Romero, G.~A., et al.\ 2012, \apj, 750, 157

\bibitem[Dong et al.(2017)]{bib:dong2017}
Dong, R., van der Marel, N., Hashimoto, J., et al.\ 2017, \apj, 836, 201

\bibitem[Dubrulle et al.(1995)]{bib:dubrulle1995} Dubrulle, B., Morfill, G., \& Sterzik, M.\ 1995, \icarus, 114, 237

\bibitem[Favre et al.(2013)]{bib:favre2013}
Favre, C., Cleeves, L.~I., Bergin, E.~A., Qi, C., \& Blake, G.~A.\ 2013, \apjl, 776, L38

\bibitem[Fedele et al.(2017)]{bib:fedele2017}
Fedele, D., Carney, M., Hogerheijde, M.~R., et al.\ 2017, \aap, 600, A72 

\bibitem[Fukagawa et al.(2013)]{bib:fukagawa2013}
Fukagawa, M., Tsukagoshi, T., Momose, M., et al.\ 2013, \pasj, 65, L14

\bibitem[de Gregorio-Monsalvo et al.(2013)]{bib:gregorio-mosalvo2013}
de Gregorio-Monsalvo, I., M{\'e}nard, F., Dent, W., et al.\ 2013, \aap, 557, A133

\bibitem[Guilloteau et al.(2016)]{bib:guilloteau2016}
Guilloteau, S., Pi{\'e}tu, V., Chapillon, E., et al.\ 2016, \aap, 586, L1

\bibitem[Hayashi(1981)]{bib:hayashi1981}
Hayashi, C.\ 1981, Progress of Theoretical Physics Supplement, 70, 35 


\bibitem[Kokubo \& Ida(2002)]{bib:kokubo2002}
Kokubo, E., \& Ida, S.\ 2002, \apj, 581, 666

\bibitem[Lada \& Wilking(1984)]{bib:lada1984}
Lada, C.~J., \& Wilking, B.~A.\ 1984, \apj, 287, 610

\bibitem[Lada(1987)]{bib:lada1987}
Lada, C.~J.\ 1987, Star Forming Regions, 115, 1


\bibitem[van der Marel et al.(2013)]{bib:marel2013}
van der Marel, N., van Dishoeck, E.~F., Bruderer, S., et al.\ 2013, Science, 340, 1199

\bibitem[van der Marel et al.(2015)]{bib:marel2015}
van der Marel, N., van Dishoeck, E.~F., Bruderer, S., P{\'e}rez, L., \& Isella, A.\ 2015, \aap, 579, A106

\bibitem[van der Marel et al.(2016)]{bib:marel2016}
van der Marel, N., van Dishoeck, E.~F., Bruderer, S., et al.\ 2016, \aap, 585, A58

\bibitem[McMullin et al.(2007)]{bib:mcmullin2007}
McMullin, J.~P., Waters, B., Schiebel, D., Young, W., \& Golap, K.\ 2007, Astronomical Data Analysis Software and Systems XVI, 376, 127 

\bibitem[Muto et al.(2015)]{bib:muto2015}
Muto, T., Tsukagoshi, T., Momose, M., et al.\ 2015, \pasj, 67, 122

\bibitem[Nomura \& Millar(2005)]{bib:nomura2005}
Nomura, H., \& Millar, T.~J.\ 2005, \aap, 438, 923

\bibitem[P{\'e}rez et al.(2014)]{bib:perez2014}
P{\'e}rez, L.~M., Isella, A., Carpenter, J.~M., \& Chandler, C.~J.\ 2014, \apjl, 783, L13

\bibitem[Pinilla et al.(2012)]{bib:pinilla2012}
Pinilla, P., Benisty, M., \& Birnstiel, T.\ 2012, \aap, 545, A81

\bibitem[Pinilla et al.(2017)]{bib:pinilla2017}
Pinilla, P., P{\'e}rez, L.~M., Andrews, S., et al.\ 2017, \apj, 839, 99

\bibitem[Pinte et al.(2008)]{bib:pinte2008}
Pinte, C., Padgett, D.~L., M{\'e}nard, F., et al.\ 2008, \aap, 489, 633

\bibitem[Pinte et al.(2017)]{bib:pinte2017}
Pinte, C., Menard, F., Duchene, G., et al.\ 2017, arXiv:1710.06450

\bibitem[van der Plas et al.(2017a)]{bib:plas2017a}
van der Plas, G., Wright, C.~M., M{\'e}nard, F., et al.\ 2017, \aap, 597, A32 

\bibitem[van der Plas et al.(2017b)]{bib:plas2017b}
van der Plas, G., M{\'e}nard, F., Canovas, H., et al.\ 2017, \aap, 607, A55 

\bibitem[Qi et al.(2011)]{bib:qi2011}
Qi, C., D'Alessio, P., {\"O}berg, K.~I., et al.\ 2011, \apj, 740, 84

\deleted{\bibitem[Pinte et al.(2016)]{bib:pinte2016}
Pinte, C., Dent, W.~R.~F., M{\'e}nard, F., et al.\ 2016, \apj, 816, 25}

\bibitem[Rice et al.(2006)]{bib:rice2006}
Rice, W.~K.~M., Armitage, P.~J., Wood, K., \& Lodato, G.\ 2006, \mnras, 373, 1619

\bibitem[Romero et al.(2012)]{bib:romero2012}
Romero, G.~A., Schreiber, M.~R., Cieza, L.~A., et al.\ 2012, \apj, 749, 79

\bibitem[Rosenfeld et al.(2013)]{bib:rosenfeld2013}
Rosenfeld, K.~A., Andrews, S.~M., Hughes, A.~M., Wilner, D.~J., \& Qi, C.\ 2013, \apj, 774, 16 

\bibitem[Salter et al.(2011)]{bib:salter2011}
Salter, D.~M., Hogerheijde, M.~R., van der Burg, R.~F.~J., Kristensen, L.~E., \& Brinch, C.\ 2011, \aap, 536, A80

\bibitem[Shakura \& Sunyaev(1973)]{bib:shakura1973}
Shakura, N.~I., \& Sunyaev, R.~A.\ 1973, \aap, 24, 337

\bibitem[Siess et al.(2000)]{bib:siess2000}
Siess, L., Dufour, E., \& Forestini, M.\ 2000, \aap, 358, 593 

\bibitem[Strom et al.(1989)]{bib:strom1989}
Strom, K.~M., Strom, S.~E., Edwards, S., Cabrit, S., \& Skrutskie, M.~F.\ 1989, \aj, 97, 1451

\bibitem[Tachihara et al.(1996)]{bib:tachihara1996}
Tachihara, K., Dobashi, K., Mizuno, A., Ogawa, H., \& Fukui, Y.\ 1996, \pasj, 48, 489

\bibitem[Tsukagoshi et al.(2014)]{bib:tsukagoshi2014}
Tsukagoshi, T., Momose, M., Hashimoto, J., et al.\ 2014, \apj, 783, 90

\bibitem[Tsukagoshi et al.(2016)]{bib:tsukagoshi2016}
Tsukagoshi, T., Nomura, H., Muto, T., et al.\ 2016, \apjl, 829, L35

\bibitem[Zhang et al.(2014)]{bib:zhang2014}
Zhang, K., Isella, A., Carpenter, J.~M., \& Blake, G.~A.\ 2014, \apj, 791, 42

\end{thebibliography}
\end{document}